\documentclass[12pt]{article}
\usepackage{epsfig,amssymb}

\newcommand{\bq}{\begin{equation}}
\newcommand{\eq}{\end{equation}}
\newcommand{\bqa}{\begin{eqnarray}}
\newcommand{\eqa}{\end{eqnarray}}
\newcommand{\ben}{\begin{enumerate}}
\newcommand{\een}{\end{enumerate}}
\newcommand{\bc}{\begin{center}}
\newcommand{\ec}{\end{center}}
\newcommand{\bqb}{\begin{eqnarray*}}
\newcommand{\eqb}{\end{eqnarray*}}

\def\lsim{\lesssim}

\def\ie{{\it i.e. }}
\def\eg{{\it e.g. }}

\def\etal{{\it et.al. }}

\def\sw{s_W}
\def\cw{c_W}
\def\tw{t_W}
\def\swd{s^2_W}
\def\cwd{c^2_W}

\def\mwd{m_W^2}
\def\mw{m_W}
\def\mz{m_Z}
\def\mzd{m_Z^2}

\def\L{ {\cal L }}
\def\C{ {\cal C }}
\def\tchi{\tilde \chi^0}
\def\tchic{\tilde \chi}

\def\cbeta{c_\beta}

\def\sbeta{s_\beta}
\def\s2beta{s_{2 \beta}}
\def\c2beta{c_{2 \beta}}
\def\ca{c_1}
\def\cb{c_\mu}

\def\cab{c_{1\mu}}
\def\exa{e^{i \Phi_1}}
\def\exb{e^{i \Phi_\mu}}
\def\exma{e^{-i \Phi_1}}

\def\exmapb{e^{-i (\Phi_1+\Phi_\mu)}}
\def\exbma{e^{i (\Phi_\mu-\Phi_1)}}
\def\lame{\lambda_1}
\def\lamp{\lambda_2}

\def\pr#1#2#3{ Phys. Rev. ${\bf{#1}}$: #2 (#3)}

\def\pl#1#2#3{ Phys. Lett. ${\bf{#1}}$: #2 (#3)}
\def\prep#1#2#3{ Phys. Rep. ${\bf{#1}}$: #2 (#3)}

\def\np#1#2#3{ Nucl. Phys. ${\bf{#1}}$: #2 (#3)}
\def\zp#1#2#3{ Z. f. Phys. ${\bf{#1}}$: #2 (#3)}
\def\epj#1#2#3{ Eur. Phys. J. ${\bf{#1}}$: #2 (#3)}

\begin{document}
\pagenumbering{arabic} \thispagestyle{empty}
\def\thefootnote{\fnsymbol{footnote}}
\setcounter{footnote}{1}

\begin{flushright}
THES-TP 2002/01
\\ hep-ph/0204152 \\
April  2002
\\
 \end{flushright}
\vspace{2cm}
\begin{center}
{\Large\bf  The neutralino projector formalism
for complex SUSY parameters\footnote{Partially
supported by EU contracts
HPMF-CT-1999-00363 and   HPRN-CT-2000-00149, and the
PLATON French-Greek Collaboration project, 2002.}.}
 \vspace{1.5cm}  \\
{\large G.J. Gounaris, C. Le Mou\"{e}l} \\ \vspace{0.4cm}
Department of Theoretical Physics, Aristotle University of
Thessaloniki,\\ Gr-541 24, Thessaloniki, Greece.\\

\vspace{2.cm}

{\bf Abstract}
\end{center}

We present a new formalism  describing the neutralino
physics in the context of the minimal
 supersymmetric model (MSSM), where
 CP violation induced by
  complex $M_1$ and $\mu$ parameters is  allowed.
  The formalism is based on the
construction of neutralino projectors, and can be directly
generalized to non-minimal SUSY models involving any number of
neutralinos. It extends a previous work
applied to the real SUSY parameter case.
 In  MSSM, the method allows to describe
 all  physical observables related to a
 specific neutralino, in terms of its
 CP eigenphase and three complex numbers called its  "reduced
 projector elements". As an example, $\sigma(e^- e^+ \to
 \tchi_i \tchi_j)$ is presented in this language.

As the experimental knowledge on the neutralino-chargino
sectors is being accumulated, the problem of
extracting the various SUSY parameters will arise.
Motivated by this, we consider various scenarios
concerning   the quantities that could be first measured.
Analytical disentangled expressions determining
the related  SUSY parameters from them, are then derived,
which  emphasize  the efficiency of the formalism.\\

\noindent
PACS number(s): 12.60.Jv

\def\thefootnote{\arabic{footnote}}
\setcounter{footnote}{0} \clearpage

\section{Introduction}

The  neutralinos are probably among the lightest
supersymmetric particles \cite{SUSY-rev, lectures}. In fact,
if R-parity is conserved,  the lightest neutralino  is
one of the most promising candidates for the
Dark Matter of the Universe \cite{Ellis-bench, Snowmass}.

On the other hand, the description of neutralinos
is quite complicated because,
even in the minimal supersymmetric model (MSSM),
   it   involves the diagonalization of a
$4 \times 4$ matrix, which determines the neutralino masses and
mixings, and thereby the various couplings \cite{SUSY-rev, lectures}.
It is also quite possible
that the true theory actually contains
 more than four neutralinos \cite{Espinosa},
 which will further complicate the situation. Such thoughts motivate
  the idea to look for a formalism which will
 simplify the neutralino problem  and, if possible,
be easy to generalize  to any    neutralino number.

In case  the Higgs mixing parameter $\mu$ and the soft breaking ones
$M_1$ and $M_2$ are all  real,  the neutralino mass-matrix
is real and symmetric, implying that the neutralino and
chargino sectors are both CP conserving.  The diagonalization
of the real neutralino mass matrix in MSSM  has been
analytically  studied since a long time
\cite{neutralinos1a, neutralinos1b, Cairo}.

Nevertheless in \cite{GLMP}, inspired by Jarlskog's treatment
of the CKM matrix \cite{Jarlskog},  we have added to the previous
descriptions a new one,   which describes each physical
neutralino in terms of its projector matrix and its
CP eigensign $\eta_j=\pm 1$. Physically, the projector matrix of
a mass-eigenstate neutralino is identified with
its density matrix in the space
of the neutral gaugino and higgsino fields.
If the neutralinos were  Dirac particles,
physical observables would only depend on these
projectors; so that  an explicit dependence on the CP eigensigns,
signals  contributions generated by the Majorana
nature of the neutralinos  \cite{GLMP}.  Analytic expressions
for these CP eigensigns and projector matrices,
as well as  the physical masses, in terms of the real
SUSY parameters $M_1$, $\mu$ and $M_2$
were  given in \cite{GLMP}.

If  $M_1$ and $\mu$ are complex,  with non-trivial phases
$\Phi_1$ and $\Phi_\mu$ respectively, then
CP  is violated in the neutralino and chargino sectors\footnote{$M_2$
is chosen by convention to be real and positive. Recent
analyses of the electric dipole moments
suggest that  $\Phi_\mu$ is probably
very close to zero or $\pi$,
provided  the sfermions of the first two generations
are not too   heavy   \cite{EDMne}.}.
Diagonalizing the neutralino mass matrix
is then more complicated; but nevertheless, explicit
analytic solutions  have already been constructed in \cite{Zerwas1}.

The first purpose of the present work is to extend the neutralino
 projector formalism of \cite{GLMP}, to the
case of complex SUSY parameters.
The formalism is applied  to the MSSM case containing
four neutralinos  with complex (or real) couplings; but it
can be  straightforwardly  extended  to  any neutralino number.

For the description of each physical neutralino we  need,
in addition to its  projector matrix,
 one pseudo-projector matrix and one CP eigenphase.
As in the real parameter  case, if the neutralinos were Dirac
particles, only the projectors would be needed.
It is their Majorana
nature that necessitates   the introduction of the
 pseudo-projectors and CP-eigenphases. This is done in  Section 2;
while in Appendix A we present (for completeness) the
formulae for calculating the neutralino  physical masses in terms
of the  MSSM parameters  at the tree level.

In Section 3  we introduce  the
notion of the "reduced  projector elements" (RPE), which
in MSSM constitute
 12 complex numbers; three for each physical neutralino.
The 12 RPE are not independent though; since the three
RPE referring to just one physical neutralino, are
sufficient to determine all the other ones.
All projector, RPE and  pseudo-projector matrix elements,
as well as the    CP-eigenphases,
are expressed in terms of the neutralino mass matrix,
in a way which can be immediately generalized
to any  neutralino number.
 Explicit expressions
for the RPE in terms of the MSSM  parameters at tree level,
are also given.
In the same Section 3, a new form
of the necessary  and sufficient
condition for CP conservation in the neutralino-chargino
sector, is also  presented.

The RPE and the neutralino CP eigenphases offer an
elegant way to describe  the  neutralino physical observables.
As an example, the cross section
for $e^-e^+\to \tchi_i \tchi_j$ with
longitudinally polarized beams, appears   in Section 4.

\vspace{0.2cm}
For the physical applications anticipated immediately after the
first charginos and/or  neutralinos will be discovered,
it is actually not sufficient to simply express
their masses and mixings in terms of the SUSY parameters
at the  electroweak scale. Inverse relations should  also be
provided,  expressing the   SUSY parameters in terms of
the partial  ino  information which might become available.
These relations depend of course  on scenarios
 about  such discoveries; and since the input  information
will be rather lacking, they will inevitably
involve  ambiguities,
whose lifting  calls for appropriate ideas.

Work in this spirit, has already appeared in \cite{Bartl3, Abdel2}.
Concerning particularly  the neutralino sector,
very detail work has been
presented in \cite{Kneur-CPc} for the CP-conserving case, and in
\cite{Kneur-CPv, Zerwas1} for the CP-violating one.
One kind of scenarios considered in these references, was  based on the
idea that, when the neutralinos will start being studied,
the chargino parameters $M_2$, $|\mu|$,
$\Phi_\mu$, and $\tan\beta$ will be  already known from
 $\tchic_j^\pm$ production experiments  and  studies of
$e^-e^+\to \tchic_i^- \tchic_j^+$ \cite{Abdel-CPv1}.
Under these conditions,  the  only (generally complex)
parameter to be  determined from the neutralino studies is
$M_1$. Explicit formulae for  determining  $M_1$, and suggestions
on the  neutralino information needed
to lift the inherent ambiguities,  have
appeared in  \cite{Kneur-CPc, Kneur-CPv, Zerwas1}.

Another kind of scenarios  was based on the idea that
  the chargino sector will  not be  fully known
 by the time  the  first neutralinos will start being studied.
 As an example, it was   assumed that
 only the lightest chargino and its mixing angles are determined
 from Linear Collider (LC), together
 with the masses of one or two of
 the lightest neutralinos. Then, relations for determining the
 generally complex $M_1$ and $\mu$ parameters
 were suggested, in which information from
 $e^-e^+\to \tchi_1 \tchi_2$ was  used  \cite{Zerwas1}.

The  second aim of this paper is
to  present the formalism for extracting the  SUSY
parameters, in the context  of   scenarios  concerning
the neutralino and chargino  measurements. We  essentially
study the same scenarios
as in  \cite{Kneur-CPc, Kneur-CPv, Zerwas1}.
This is done in Section 5.
In all cases, disentangled expressions are given, determining
 the relevant  SUSY parameters from the input data, and many
 numerical examples are presented.
They are   based  on a  recently proposed set of benchmark
SUGRA models \cite{Snowmass};  to which variations are made,
 in order for $M_1$ and $\mu$ to become complex.
Scenarios involving  real   SUSY parameters, are also
considered, based on  the SUGRA benchmarks
 of \cite{Snowmass, Ellis-bench}.
It is checked  that the relations constructed
in the various
scenarios, really determine the SUSY parameters they are supposed
to. Our treatment is  theoretical though, simply aiming
at supplying the relevant formulae. No error in the input data  of
the various scenarios, is taken into consideration.

The whole analysis is carried at tree level,
but in principle it could be extended
  to any order of perturbation theory, essentially by simply using
the appropriate renormalized neutralino mass matrix.
In such a case of course, parameters from other SUSY sectors will
enter, leading to the necessity of a more global analysis \cite{Hollik}.

In Section 6 we give our conclusions. Finally,
in Appendix B we compare to the approach of
 \cite{Zerwas1} for MSSM,  and identify the relations  between the
elements  of the two   formalisms.

\section{The neutralino projectors, pseudo-projectors
and CP eigenphases.}

A  description of neutralinos in terms of their
projectors has already been proposed in \cite{GLMP}, where
the restriction to real $M_1, ~M_2, ~\mu$  parameters was made,
corresponding to CP conserving  chargino and neutralino
sectors.

In the present work,  we  extend
this formalism to  the most general  CP-violating case in which
$M_1$ and $\mu$ are taken  complex, \ie
\bq
M_1 \equiv \exa \bar M_1 ~~~~, ~~~~  \mu \equiv \exb \bar \mu ~~~,
\label{M1-mu-def}
\eq
where $\bar M_1 \equiv |M_1|$,  $\bar \mu \equiv|\mu|$,  while
the phase angles are defined such that  $-\pi <\Phi_1 \leq \pi$ and
$-\pi <\Phi_\mu \leq \pi$. Without loss of generality,
the soft SUSY breaking gaugino
mass $M_2$, is always taken real and positive.

The neutralino mass term in the  SUSY
Lagrangian is written as  \cite{SUSY-rev}
\bq
\L_m= ~-\frac{1}{2} \Psi^{0\top }_L \C Y \Psi^0_L ~+~ {\rm
h.c.} ~~ , ~~~~ \label{mass-term}
\eq
where the minimal complex symmetric neutralino mass-matrix is
\bq
Y=  \left ( \matrix{\bar M_1 e^{i \Phi_1} & 0  & -\mz \sw \cbeta
& \mz \sw \sbeta \cr 0 & M_2 & \mz \cw \cbeta & -\mz \cw \sbeta
\cr -\mz \sw \cbeta& \mz \cw \cbeta & 0 & - \bar \mu e^{i \Phi_\mu}\cr
\mz \sw \sbeta & -\mz \cw \sbeta & -\bar \mu e^{i \Phi_\mu} & 0
\cr}\right ) ~~ ,
 \label{Y-matrix}
 \eq
 $\C =i\gamma^2 \gamma^0$  is the usual
Dirac charge conjugation matrix, and  $\Psi^0_L$
is the column vector describing  the Left
neutralino fields  in the "weak basis"
of the gauginos and Higgsinos   as
\bq
\Psi^0_L \equiv \left ( \matrix{\tilde B_L \cr \tilde
W^{(3)}_L \cr \tilde H_{1L}^0 \cr  \tilde H_{2L}^0 \cr } \right )
~~ .  \label{Psi0L}
\eq

The corresponding Left  mass eigenstate fields
 $\tchi_{jL}$ are related to them through the
  unitary transformation $U_N$
as
\bq
\Psi^0_{\alpha L}=\sum_{j=1}^4 U_{N\alpha j}\tchi_{jL} ~~,
\label{UN-definition}
\eq
with $(\alpha,~j= 1~-~4)$ being   respectively the  indices
for the  weak and mass  eigenstates. The matrix $Y$
is   diagonalized as
\bq
U_N^\top Y U_N =\left (  \matrix{ m_{\tchi_1} & 0 & 0& 0 \cr
 0 & m_{\tchi_2} & 0 & 0 \cr
 0 & 0 & m_{\tchi_3} & 0 \cr
 0 & 0 & 0&  m_{\tchi_4}  \cr }\right ) ~~
, ~~ \label{UN-Yd-matrix}
\eq
where   $m_{\tchi_j}~ (j=1-4)$ describe the
physical  neutralino masses ordered as
\bq
0\leq  m_{\tchi_1} \leq m_{\tchi_2} \leq m_{\tchi_3} \leq
m_{\tchi_4} ~~ . \label{mj-ordering}
\eq
Using (\ref{UN-Yd-matrix}) we get
\bqa
&& U_N^\top Y U_N = \sum_{j=1}^4 m_{\tchi_j} E_j ~~,
\label{Y-analysis} \\
&& U_N^\dagger Y^\dagger Y U_N = \sum_{j=1}^4 m_{\tchi_j}^2 E_j ~~,
\label{YdagY-analysis}
\eqa
where $E_j$ are the basic  $4\times 4$  matrices defined by
$(E_j)_{ik}=\delta_{ji}\delta_{jk}$.

\vspace{0.2cm}
The neutralino projector
matrices, which acting on the weak-basis fields, project out
the neutralino mass eigenstate $\tchi_j$, are  defined by \cite{GLMP}
\bq
P_j =P_j^\dagger=  U_N E_j U_N^\dagger ~~, \label{Pj-def}
\eq
so that
\bq
P_{j\alpha \beta}=U_{N\alpha j}U_{N\beta j}^* ~~.
\label{Pj-elements}
\eq
They   satisfy  the standard  projector relations
\bqa
&& P_i P_j=P_j \delta_{ij} ~~,~~ TrP_j=1 ~~ ,~~\sum_{j=1}^4 P_j=1 ~~,
\label{Pj-property1} \\
&& P_{j\alpha \alpha}P_{j\beta \beta}=|P_{j\alpha\beta}|^2
~~ , \label{Pj-property2}
\eqa
where summation over the repeated indices is only done
if it is explicitly indicated.
Here  $(i,~j)=(1-4)$ describe the neutralino mass-eigenstate
indices, while
$(\alpha, ~\beta)=(1-4)$ are the weak basis ones.
Eq.(\ref{Pj-property2}) which follows from
(\ref{Pj-elements}),  may also be obtained by
viewing the  projector
as the  density matrix of  a pure   state  \cite{Landau}. From
(\ref{YdagY-analysis}, \ref{Pj-def}) we also get
\bq
  Y^{\dagger}Y=\sum_{j=1}^4 m_{\tchi_j}^2 P_j ~~ .
\label{YdagY-analysis1}
\eq

The analytic expressions for the physical
 neutralino masses are obtained by
solving the characteristic equation for $Y^\dagger Y$,
which at tree level is given by
(\ref{physical-masses}) in Appendix A. They are ordered
according to  (\ref{mj-ordering}).
Following Jarlskog \cite{Jarlskog}, the neutralino projectors
are then given by \cite{GLMP}
\bqa
P_1 &=& \frac{( m_{\tchi_4}^2 -Y^\dagger Y) ( m_{\tchi_3}^2
-Y^\dagger Y)(  m_{\tchi_2}^2 -Y^\dagger Y)} {(  m_{\tchi_4}^2 -
m_{\tchi_1}^2) (  m_{\tchi_3}^2 -  m_{\tchi_1}^2) ( m_{\tchi_2}^2
-  m_{\tchi_1}^2)} ~~ ~~, \nonumber \\[0.3cm] P_2 &=& \frac{(
m_{\tchi_4}^2 -Y^\dagger Y) (  m_{\tchi_3}^2 -Y^\dagger Y)
(Y^\dagger Y-
  m_{\tchi_1}^2 )} {(  m_{\tchi_4}^2 -  m_{\tchi_2}^2)
(  m_{\tchi_3}^2 -  m_{\tchi_2}^2) (  m_{\tchi_2}^2 -
m_{\tchi_1}^2)} ~~ ~~, \nonumber \\[0.3cm] P_3 &=& \frac{(
m_{\tchi_4}^2 -Y^\dagger Y) (Y^\dagger Y-
  m_{\tchi_2}^2 )(Y^\dagger Y-
m_{\tchi_1}^2 )} {(  m_{\tchi_4}^2 -  m_{\tchi_3}^2) (
m_{\tchi_3}^2 -  m_{\tchi_2}^2) (  m_{\tchi_3}^2 - m_{\tchi_1}^2)}
~~ ~~, \nonumber \\[0.3cm] P_4 &=& \frac{(Y^\dagger Y-
m_{\tchi_3}^2 ) (Y^\dagger Y-   m_{\tchi_2}^2 )(Y^\dagger Y-
m_{\tchi_1}^2 )} {(  m_{\tchi_4}^2 -  m_{\tchi_3}^2) (
m_{\tchi_4}^2 -  m_{\tchi_2}^2) (  m_{\tchi_4}^2 -
m_{\tchi_1}^2)} ~~ ~~, \label{Pj-Jarlskog}
\eqa
where (at tree level again) $Y^\dagger Y$ is taken from
 (\ref{YdagY-matrix}). In (\ref{Pj-Jarlskog})  the
assumption is made, that all neutralino masses are different
from each other\footnote{In case of mass degeneracy,
the instructions  in \cite{Jarlskog} should be followed.},
and that there are only four neutralinos.
The extension of this formalism to cases involving
any number of  neutralinos and any form of the
$Y$-matrix\footnote{For more than four neutralinos,
their  physical masses
may be obtained by solving \eg numerically,
the characteristic equation for $Y^\dagger Y$.} is obvious
\cite{Jarlskog}.   \par

\vspace{0.2cm}
As already mentioned, if the neutralinos were Dirac particles,
 the projectors in (\ref{Pj-Jarlskog}),
 would had been sufficient to describe any physical observable.
Because of the Majorana nature of the  neutralinos though,
additional parameters are needed,  denoted as
their  CP eigenphases and pseudo-projector matrices.

The CP-eigenphases, are analogous to
the CP eigensigns of the real SUSY-parameter
case treated in \cite{GLMP}, and are defined as
(compare (\ref{UN-definition}))
\bq
\eta_j \equiv \frac{(U_N^*)_{1j}} {(U_N)_{1j}} ~~,
\label{etaj-def}
\eq
where $|\eta_j|= 1$ is obvious. The  phases
in (\ref{etaj-def}),  chosen as\footnote{ The CP conserving
case  is  obtained when  ${\rm Arg}(\eta_j)$ is either  0 or  $+\pi$,
\cite{GLMP}.}
\[
- \pi <  {\rm Arg} (\eta_j) \leq \pi ~~ ,
\]
are  determined by those of the first row of $U_N$.
Using  $\eta_j$, a second unitary matrix $U^0$ is defined
through \cite{GLMP}
\bq
(U_N)_{\alpha j} \equiv \sqrt{\eta_j^*} ~ U^0_{\alpha j} ~~,
\label{U0-def}
\eq
so that  (\ref{UN-definition}) may be rewritten as
\bq
 \Psi^0_{\alpha L}= \sum_{j=1}^4 U_{N \alpha j}\tchi_{jL}
= \sum_{j=1}^4 \sqrt{\eta_j^*} ~ U_{\alpha j}^0\tchi_{jL} ~~~.
\label{U0UN-matrix}
\eq
Using  (\ref{U0-def}, \ref{Pj-elements}), the $P_j$ matrix
elements may  be expressed as
\bq
P_{j\alpha\beta}= U_{N\alpha j}U_{N \beta j}^*=
 U_{\alpha j}^{0}U_{\beta j}^{0*} ~~,
\label{Pj-elements1}
\eq
which clearly indicates that the projectors  do not
know about the CP-eigenphases.

The use of (\ref{UN-Yd-matrix}, \ref{U0-def}) also  give
\bq
 U^{0\top} Y U^0 =
 \left (  \matrix{\tilde  m_{\tchi_1} & 0 & 0& 0 \cr
 0 &\tilde  m_{\tchi_2} & 0 & 0 \cr
 0 & 0 &\tilde  m_{\tchi_3} & 0 \cr
 0 & 0 & 0& \tilde  m_{\tchi_4}  \cr }\right ) ~~
, ~~ \label{U0-Yd-matrix}
\eq
where the  "complex" neutralino masses
$\tilde m_{\tchi_j}~ (j=1-4)$ are
related to the physical (positive) ones $m_{\tchi_j}$ by
\bq
\tilde m_{\tchi_j}  \equiv \eta_j m_{\tchi_j}~~~~,
\label{complex-masses}
\eq
and constitute a simple generalization of the "signed" masses
used in  the real SUSY parameter case \cite{GLMP}.
In addition,  (\ref{U0-Yd-matrix}, \ref{Y-analysis}) imply
\bq
 U^{0\top} Y U^0 = \sum_{j=1}^4 \tilde m_{\tchi_j} E_j ~~,
\label{Y-U0-analysis}
\eq
which is analogous to (\ref{Y-analysis}), but
involving the complex masses
instead\footnote{Eqs.(\ref{U0UN-matrix}, \ref{complex-masses})
look  formally the same as Eqs.(7, 6) of \cite{GLMP}, in which
the SUSY parameters were assumed real. In the present formalism,
the quantity $\tilde \eta_j$ of \cite{GLMP} should be identified as
$\tilde \eta_j =\sqrt{\eta_j^*}$.}.

\vspace{0.2cm}
We next turn to defining
the pseudo-projector  for the  physical
neutralino $\tchi_j$ as
\bq
\bar P_j=\bar P_j^\top = U_N^* E_j U_N^\dagger
= \eta_j U^{0*} E_j U^{0\dagger} ~~,
\label{Pj-bar-def}
\eq
with its matrix elements being
\bq
\bar P_{j\alpha \beta}=U_{N\alpha j}^*U_{N\beta j}^*=
\eta_j U_{\alpha j}^{0*}U_{\beta j}^{0*} ~~,
\label{Pj-bar-elements}
\eq
and  satisfying
\bqa
&& \bar P_j^* \bar P_k =\delta_{jk} P_j ~~,
\label{Pj-bar-properties1}  \\
&& \bar P_j P_k= P_k^\top \bar P_j= \delta_{jk} \bar P_j
~~, \label{Pj-bar-properties2}
\eqa
\bq
Y=\sum_{j=1}^4 m_{\tchi_j} \bar P_j ~~.  \label{rec-Y}
\eq

As indicated by (\ref{Pj-property1}, \ref{Pj-bar-properties1},
 \ref{Pj-bar-properties2}) and
(\ref{YdagY-analysis1}, \ref{rec-Y}), the
pseudo-projector $\bar P_j$, shares the
projector $P_j$ property to isolate the $\tchi_j$ component, when
acting on any neutralino state. This is the reason we call it
pseudo-projector.

According to (\ref{Pj-bar-elements}),
 the pseudo-projector matrix elements (contrary to the
projector ones), do depend on the CP-eigenphases.
In fact, (\ref{Pj-elements1},
\ref{Pj-bar-elements})   indicate that  the
matrix elements of $P_j$ and $\bar P_j$, just differ  by
 phases, which in the CP-conserving limit
(corresponding to real  $U^0$), are simply reduced to an $\eta_j=\pm
1$ overall  factor.

\section{The reduced  projector elements.}

The neutralino  description
in terms of projectors, pseudo-projectors and CP-eigenphases
is, of course complete, but  uneconomical.
Moreover, we have not yet given the formulae from which the
CP-eigenphases and the pseudo-projector matrix elements may be
calculated. To cure this, we introduce below  the notion of the
reduced projector elements.

We first count the number of independent real parameters
needed for a complete description. This
 is  determined by the number of parameters
entering the neutralino mass matrix, which at tree level has the form
appearing in  (\ref{Y-matrix}).
If we temporarily ignore its  specific form,
and consider it as a general  complex
symmetric $4 \times 4$ matrix, there would  be
  20   independent parameters. As such, we could take
   the four physical  neutralino masses, the four
   CP eigenphases $\eta_j$,
    and the    12 complex (but constrained\footnote{For the
    counting to be correct, $p_{j\alpha}$ have to satisfy 12 real
    constraints discussed below.} )
    parameters defined as
\bq
\label{pj-variables}
p_{j\alpha}\equiv \frac{P_{j1 \alpha}}{P_{j11}}~~ ,
\eq
with $(j=1-4)$ and $(\alpha=2,3,4)$.
In writing (\ref{pj-variables}), we assume
$P_{j11}\neq 0$, which means that there should  always be
 some non-vanishing  Bino contribution to the physical
 neutralino states \cite{Cairo}. We also
define $p_{j1}=1$, so that (\ref{pj-variables}) may be extended  to
$\alpha=1$.

The quantities $p_{j\alpha}$, are directly calculated from
the Jarlskog relation (\ref{Pj-Jarlskog}), and play
a central role in the analysis below. We call them
"reduced  projector elements".
In terms of them,   the  elements of
$P_j$, $\bar P_j$ and $U_N$  are expressed by the Ansatz
\bqa
P_{j\alpha \beta} & = & P_{j11}~ p_{j \alpha}^* ~ p_{j\beta}
~~,  \label{P-pj}  \\
\bar P_{j \alpha \beta} &=& P_{j11}~ \eta_j~ p_{j \alpha}~ p_{j \beta}
~~ \label{Pbar-pj}  \\
U_{N\alpha j}&=& \sqrt{\eta_j^* P_{j11}}~ p_{j\alpha}^* ~~,
\label{Un-pj}
\eqa
where
\bq
P_{j11}= \frac{1}{1+ |p_{j2}|^2+|p_{j3}|^2+|p_{j4}|^2} ~~.
\label{Pj11-pj}
\eq
We  note that the CP-eigenphases entering
(\ref{Pbar-pj}, \ref{Un-pj}),
are the only quantities which are still needed in order to
fully determine the neutralino properties, in
terms of the initial SUSY parameters.
They will be determined  below  using (\ref{rec-Y}).

The expressions (\ref{P-pj} - \ref{Pj11-pj}), together with
 (\ref{pj-variables}, \ref{etaj-def}),
 guarantee that all projector and
pseudo-projector properties
(\ref{Pj-elements}-\ref{Pj-property2},
\ref{Pj-bar-elements}-\ref{Pj-bar-properties2}), and
 the unitarity of   $U_N$, are automatically satisfied, for any
complex symmetric neutralino mass matrix.
It is very  important  to remark that all these expressions,
can be directly generalized to any neutralino number,
  and any complex symmetric  mass matrix.

Returning  to   the counting of the independent variables
 for a general $4\times 4$ matrix $Y$, we
remark that the unitarity of $U_N$ together with
(\ref{Un-pj}) produce 12 real constraints on the
 12 complex parameters $p_{j2},p_{j3},p_{j4} ~(j=1,4)$.
Thus,   the four neutralino masses, the four
 CP eigenphases $\eta_j$, and  the 12 complex  $p_{j\alpha}$,
 indeed provide 20 independent real parameters, matching those
 contained in a most general complex symmetric, $4\times 4$,
 neutralino mass matrix, as mentioned above.

\vspace{0.2cm}
At the tree level, in which (\ref{Y-matrix}) is used,
the number of independent parameters is
 reduced from 20 to the six basic SUSY parameters
 $\bar  M_1 \exa$, $M_2$, $\bar \mu \exb$ and $\tan\beta$. This allows
to express all $p_{j\alpha}$ in terms of them.
Indeed using (\ref{Pj-Jarlskog}, \ref{pj-variables})
we find ($j=1-4$, $\alpha=1-4$)
\bq
p_{j \alpha}=\frac{q_{j \alpha}}{q_{j1}} ~~,
\label{pj-variables1}
\eq
with
\bqa
q_{j1}&=&- \mzd [ -\mzd (m_{\tchi_j}^2-\bar \mu^2
\s2beta^2) (m_{\tchi_j}^2 \swd-\bar  M_1^2 \cw^4-M_2^2 \sw^4-2
\bar  M_1 M_2 \swd \cwd \ca) \nonumber \\
&&+(m_{\tchi_j}^2-\bar \mu^2) [-2 \bar  M_1
\bar \mu \ \s2beta (\bar  M_1 M_2 \cwd \cb-
(m_{\tchi_j}^2-M_2^2) \swd \cab) \nonumber \\
&& + m_{\tchi_j}^2
((m_{\tchi_j}^2-M_2^2) \swd-2 \bar  M_1^2 \cwd)]]
 -\bar  M_1^2 (m_{\tchi_j}^2-M_2^2) (m_{\tchi_j}^2-\bar \mu^2)^2  ~,
 \nonumber \\
q_{j2}&=&- m_{\tchi_j}^2 \mzd \sw \cw [\mzd
(m_{\tchi_j}^2-\bar \mu^2 \s2beta^2)\nonumber \\
 && ~~~-(   m_{\tchi_j}^2-\bar \mu^2) (m_{\tchi_j}^2+
 \bar  M_1 M_2 \exma+
  \bar \mu \ \s2beta (\bar  M_1 \exmapb+M_2 \exb))] ~~, \nonumber \\
q_{j3}&=&m_{\tchi_j}^2 \mz \sw [-\mzd
\bar \mu \sbeta \c2beta (\bar  M_1 M_2 \cwd \exbma-
(m_{\tchi_j}^2-M_2^2 \swd) \exb) \nonumber \\ && ~~~~~~~~+\mzd
\cwd \cbeta (m_{\tchi_j}^2-2 \bar \mu^2 \sbeta^2) (M_2-\bar  M_1 \exma)
\nonumber \\ && ~~~~~~~~+ (m_{\tchi_j}^2-\bar \mu^2)
(m_{\tchi_j}^2-M_2^2) (\bar  M_1 \cbeta \exma+\bar \mu \sbeta \exb)] ~~,
\nonumber \\
q_{j4}&=&-m_{\tchi_j}^2 \mz
\sw [\mzd \bar \mu \cbeta \c2beta (\bar  M_1 M_2 \cwd \exbma-
(m_{\tchi_j}^2-M_2^2 \swd) \exb) \nonumber \\ && ~~~~~~~~+\mzd
\cwd \sbeta (m_{\tchi_j}^2-2 \bar \mu^2 \cbeta^2) (M_2-\bar  M_1 \exma)
\nonumber \\ && ~~~~~~~~+ (m_{\tchi_j}^2-\bar \mu^2)
(m_{\tchi_j}^2-M_2^2) (\bar  M_1 \sbeta \exma+\bar \mu \cbeta \exb)]
~~, \label{qj-variables}
 \eqa
where
$\c2beta= \cos 2 \beta$ and
$\ca = \cos \Phi_1$, $\cb = \cos \Phi_\mu$,
$\cab = \cos(\Phi_1+\Phi_\mu)$.
We note that $q_{j3} \leftrightarrow -q_{j4}$, when exchanging
$\sbeta \leftrightarrow \cbeta$.
The physical neutralino masses appearing in (\ref{qj-variables}),  are
expressed in terms of the basic  SUSY
parameters using the formalism in Appendix A.

\vspace{0.2cm}
We next turn to the construction of formulae from which the CP
eigenphases $\eta_j$ may be calculated.
The pseudo-projector property (\ref{Pj-bar-properties1})
allows the  inversion of (\ref{rec-Y}), which combined with
(\ref{complex-masses},  \ref{Pbar-pj}), leads to
\bqa
\label{eqeta0}
\tilde m_{\tchi_j}=\eta_j m_{\tchi_j}
&=&\sum_{\alpha =1}^4 Y_{\alpha \beta}
\frac{p^*_{j \alpha}}{p_{j \beta}} ~~,  \\
\label{eqeta1}
&=& ~ \bar  M_1 e^{i
\Phi_1}-\mz \sw (\cbeta p_{j3}^*-\sbeta p_{j4}^*)~~, \\
\label{eqeta2}
&=& ~\frac{M_2 p_{j2}^*+\mz \cw (\cbeta p_{j3}^*-\sbeta
p_{j4}^*)}{p_{j2}} ~~, \\
\label{eqeta3}
&=& - ~ \frac{\bar \mu e^{i \Phi_\mu}
p_{j4}^*+\mz \cbeta (\sw -\cw p_{j2}^*)}{p_{j3}} ~~, \\
\label{eqeta4}
&=&- ~ \frac{\bar \mu e^{i \Phi_\mu} p_{j3}^*-\mz \sbeta (\sw -\cw
p_{j2}^*)}{p_{j4}} ~~.
\eqa
Using also  (\ref{physical-masses}), we thus get
  four equivalent relations expressing
the CP-eigenphases $\eta_j ~ (j=1-4)$, in terms of the reduced
projector elements and various SUSY parameters.

Inverting  them,
determines  the fundamental SUSY parameters
in terms of  $p_{j2}, ~p_{j3}, ~p_{j4}$, (for any fixed   $j$), as
 \bqa
\label{M1c} \bar  M_1 \exa &=& \tilde m_{\tchi_j}+ \mz \sw (\cbeta
p_{j3}^*- \sbeta p_{j4}^*) ~~, \\
\label{M2-magnitude}
M_2&=&\frac{p_{j2}}{p_{j2}^*}
\left [ \tilde m_{\tchi_j}-\frac{\mz \cw}{p_{j2}}
(\cbeta p_{j3}^*-\sbeta p_{j4}^* )\right ]~~, \\
\label{muc}
\bar \mu \exb &=& \mz \frac{(\cbeta p_{j4}+\sbeta p_{j3} ) (\sw-\cw
p_{j2}^*)}{|p_{j3}|^2-|p_{j4}|^2}~~,
\eqa
with the "complex" neutralino mass of  $\tchi_j$ given
by (compare  (\ref{complex-masses}))
\bq
\label{mchic}
\tilde m_{\tchi_j}=\eta_j m_{\tchi_j}=
-\mz \frac{( \cbeta p_{j3}^*+ \sbeta p_{j4}^*)
(\sw-\cw p_{j2}^*)}{|p_{j3}|^2-|p_{j4}|^2} ~~,
\eq
which  may also be viewed as expressing $\eta_j$ in terms of the
reduced projector elements, the physical neutralino mass,
 and $\tan\beta$.

Relations (\ref{M1c}-\ref{mchic}),
together with their  inverse
(\ref{eqeta1}-\ref{eqeta4}),   are very   important.
They  will be heavily used below in  exploring strategies for
determining the SUSY parameters, under various conditions
concerning the experimental knowledge of the neutralino and chargino
masses.

In the true    CP violating case where
 $\Phi_1, \Phi_\mu $ are non-trivial,
 the vanishing of the imaginary part of (\ref{M2-magnitude})
 also gives
 \bq
 \label{tanbeta-cp}
 \tan \beta=-\frac{\tw {\rm Im}[p_{j3} (p_{j2}^*)^2] +(|p_{j3}|^2
 -|p_{j2}|^2-|p_{j4}|^2){\rm Im}[p_{j3} p_{j2}^*]}
 {\tw {\rm Im}[p_{j4} (p_{j2}^*)^2]
+(|p_{j4}|^2-|p_{j2}|^2-|p_{j3}|^2) {\rm Im}[p_{j4} p_{j2}^*]}~~,
\eq
where
$t_W=\sw/\cw$, and ${\rm Im}[p]$ stands for the
imaginary part of the complex numbers $p_{j\alpha}$.
No such  relation is  obtained  in
the CP conserving case,  in which  $M_1$ and $\mu$ are both real.

Thus, in the CP-violating case, the 3 complex numbers $p_{j1}, p_{j2},
p_{j3}$ (for any fixed $j$) fully determine through
(\ref{M1c}-\ref{tanbeta-cp}), the 6 parameters
$(M_2, ~\bar  M_1 \exa,~ \bar \mu \exb, ~ \tan\beta)$
  entering the neutralino mass matrix $Y$ in (\ref{Y-matrix}).

\vspace{0.2cm}
The final topic in this Section
concerns  the condition for CP conservation
in the neutralino-chargino  sector; \ie
the condition for $M_1$ and $\mu$ to be real.
Such a condition has already
been presented in \cite{Zerwas1}. Here we derive a new
 form for it. From (\ref{pj-variables1}-\ref{qj-variables}),
we deduce  that CP conservation can only arise
 if the three $p_{j1}, p_{j2}, p_{j3}$
(for any fixed $j$), are all real. If this happens,
then according to
(\ref{M1c}-\ref{mchic}), $M_1$ and $\mu$ are also real;
which subsequently implies that all reduced projector elements
are in fact real; and all CP eigenphases
satisfy\footnote{In the CP conserving case
$\eta_i$ coincide with the corresponding quantities defined
in \cite{GLMP}. }  $\eta_i=\pm 1$, so that they may be called
CP eigensigns. Thus, the necessary and sufficient
condition for  CP conservation in the neutralino-chargino sector
is expressed by the equivalence
\bq
{\rm CP ~conservation} ~~~ \Leftrightarrow
~~~      {\rm Im}(p_{j\alpha})=0 ~~, \label{basic-cp}
\eq
for any fixed $j$, and all $\alpha$.

We note that the explicit formulae (\ref{qj-variables},
\ref{eqeta1}-\ref{tanbeta-cp}) have been derived at the tree level,
where the mass matrix is given by (\ref{Y-matrix}).
It should be possible though to extend the
formalism  to any order, by simply using
in (\ref{Pj-Jarlskog}, \ref{pj-variables}, \ref{eqeta0}),  the
renormalized  neutralino mass matrix. We should then find that the
CP-conservation condition (\ref{basic-cp}) remains true, even if
higher order effects are included.

It is amusing to compare the  criterion (\ref{basic-cp}),
to the one of   \cite{Zerwas1},
 where the quantities
\bqa
\label{quad-D}
D_{\alpha \beta} &\equiv& \sum_{j=1}^4 U_{N_\alpha j}
U^*_{N \beta j}= \sum_{j=1}^4
P_{j\alpha \beta}=\delta_{\alpha \beta} ~~, \\
\label{quad-M}
M_{ij} &\equiv&
\sum_{\alpha=1}^4 U_{N \alpha j} U^*_{N \alpha i}=
\sqrt{\eta_i \eta_j^*} ~\sum_{\alpha=1}^4
\frac{P_{i1\alpha}P_{j\alpha 1}}{\sqrt{P_{i11} P_{j11}}}=
\sqrt{\eta_i \eta_j^*}\delta_{ij} ~~,
\eqa
are constructed.  According to \cite{Zerwas1},
 CP conservation\footnote{See also Appendix B.},
 is equivalent to  requiring  that all
 terms in each of the indicated summations in
 (\ref{quad-D}, \ref{quad-M}), are either purely
 real or purely  imaginary. Remembering
(\ref{P-pj}- \ref{Pj11-pj}), we  see that this  is indeed
 equivalent to requiring the reality
of $p_{j\alpha}$. If $p_{j\alpha}$ are real, then
all terms in (\ref{quad-D}) will also be real, while
those in (\ref{quad-M}) will be either  real or  imaginary,
depending on whether $\eta_i \eta_j= + 1$ or $-1$.
\par

In the CP conserving  case, the  solution of the real equations
 (\ref{eqeta1} - \ref{eqeta4}) gives
\bqa
p_{j2}&=&-\frac{\tilde m_{\tchi_j}-M_1}{t_W (\tilde
m_{\tchi_j}-M_2)}  ~~, \nonumber \\
p_{j3}&=&- \frac{\mu (\tilde m_{\tchi_j}-M_1) (\tilde
m_{\tchi_j}-M_2) +\mzd \sbeta \cbeta (\tilde m_{\tchi_j}-M_1
\cwd-M_2 \swd)}{\mz \sw (\tilde
m_{\tchi_j}-M_2) (\tilde
m_{\tchi_j} \sbeta+\mu \cbeta)} ~~, \nonumber \\
p_{j4}&=& ~ \frac{\tilde m_{\tchi_j} (\tilde m_{\tchi_j}-M_1) (\tilde
m_{\tchi_j}-M_2) -\mzd \cbeta^2 (\tilde m_{\tchi_j}-M_1
\cwd-M_2 \swd)}{\mz \sw (\tilde
m_{\tchi_j}-M_2) (\tilde
m_{\tchi_j} \sbeta+\mu \cbeta)} ~~, \label{pj-variables1-CPc}
\eqa
which determine   $p_{j2}$, $p_{j3}$ and $p_{j4}$, in terms of
 $M_1, ~M_2, ~\mu$, and $\tilde m_{\tchi_j}$.
In  the CP conserving limit, we always take  $\mu$ and $M_1$ to be
 real of any sign, while $M_2>0$.

 Eqs. (\ref{pj-variables1-CPc})
are  analogous to (\ref{pj-variables1}, \ref{qj-variables}),
of the more  general CP violating   case. Contrary to them though,
 (\ref{pj-variables1-CPc}) involve the "signed"
neutralino masses\footnote{As in \cite{GLMP}, the "complex masses"
of (\ref{complex-masses}) are simply called
"signed masses" in the CP conserving case.}
$\tilde m_{\tchi_j}$, rather than the physical ones; compare
(\ref{complex-masses}). Partly because of this,
the symmetry relation between $p_{j3}$ and $p_{j4}$ observed in
(\ref{qj-variables}), is lost in (\ref{pj-variables1-CPc}).

We also note that since (\ref{tanbeta-cp}) is not obtained
in  the CP conserving case, $\tan\beta$ must be known from
some other means, in order to use the three real
($p_{j1}$, $p_{j2}$, $p_{j3}$)   for determining
$M_1$, $M_2$ and $\mu$ from (\ref{M1c}-\ref{mchic}).

\vspace{0.2cm}
Before concluding this section, we recapitulate the  procedure
for diagonalizing the neutralino mass matrix,
in the most general complex parameter case. For this, one  first
constructs the physical neutralino masses in terms of the SUSY
parameters, using the formalism in Appendix A. Then,
(\ref{Pj-Jarlskog}, \ref{pj-variables}) or
(\ref{pj-variables1}, \ref{qj-variables}), allow to construct the
projector and  reduced projector elements, from which
 the pseudo-projectors and the CP-eigenphases may be obtained
through (\ref{Pbar-pj}) and any of
(\ref{eqeta1}-\ref{eqeta4}, \ref{mchic}).
If needed, $U_N$ may then be obtained from (\ref{Un-pj}).

 The results  in the CP conserving limit, may  be obtained
by simply putting $(\Phi_1, ~\Phi_\mu) = 0$ or $\pi$, in the
aforementioned  formulae.
Alternatively, they can also be derived
using (\ref{pj-variables1-CPc})
and the  "signed" neutralino masses calculated in the
 Appendix of \cite{GLMP}.

\section{The $e^-e^+ \to \tchi_i\tchi_j$ cross section.}

The neutralino propagators to lowest
order in the weak basis, or complex SUSY parameters, are  \cite{GLMP}
\bqa
 &&
 \langle 0|T \Psi_{\alpha L}^0(x)\Psi_{\beta L}^{0\top}(y) |0
\rangle = -\sum_{j=1}^4  m_{\tchi_j}\bar P_{j\alpha \beta}^*
\Delta_F(x-y;~ m_{\tchi_j}) ~\C  ~\frac{(1-\gamma_5)}{2}
 ~~ , \label{neutralino-propag1} \\
 &&
  \langle 0|T \Psi_{\alpha L}^{0\dagger \top}(x)
\Psi_{\beta L}^{0 \dagger}(y) |0 \rangle =
\sum_{j=1}^4  m_{\tchi_j}\bar P_{j\alpha \beta}
\Delta_F(x-y;~ m_{\tchi_j}) ~\C  ~\frac{(1-\gamma_5)}{2}
 ~~ , \label{neutralino-propag2} \\
&&
\langle 0|T \Psi_{\alpha L}^{0}(x)\Psi_{\beta L}^{0 \dagger}(y)
 |0 \rangle = \sum_{j=1}^4 P_{j\alpha \beta} S_F^{(1)}(x-y; ~
m_{\tchi_j} ) \gamma_0 ~\frac{(1-\gamma_5)}{2}
 ~~ , \label{neutralino-propag3}
\eqa
and
\bqa
\Delta_F(x-y;~ m) &= & i \int \frac{d^4k}{(2\pi)^4} e^{-i
k(x-y)}\frac{1}{k^2-m^2+i\epsilon} ~~ , \nonumber \\
S_F^{(1)}(x-y; ~ m) &= & i \int \frac{d^4k}{(2\pi)^4} e^{-i
k(x-y)}\frac{\rlap /k }{k^2-m^2+i\epsilon} ~~ .
\label{propagator-functions}
\eqa

If the neutralinos were Dirac particles, only
the propagator  (\ref{neutralino-propag3}) would have been allowed,
implying  that the projector matrix elements
would have been sufficient to express all physical observables.
The Majorana nature of the neutralinos though,  introduces
the additional propagators  (\ref{neutralino-propag1},
\ref{neutralino-propag2}), and thereby  necessitates
  the introduction of the pseudo-projectors also.

Using subsequently  (\ref{P-pj}, \ref{Pbar-pj}, \ref{Pj11-pj}),  we
conclude that an elegant way to  express
 the  physical observables related to a  neutralino
$\tchi_j$, is  in terms of its three reduced projector
elements $p_{j\alpha}$ ($\alpha=2,3,4 $),  its CP eigenphase
$\eta_j$ and  its physical mass.

As an example, we consider the
differential cross section for $e^- e^+ \to \tchi_i \tchi_j$,
using  longitudinally polarized  $e^-$ and $e^+$ beams,
with the  polarizations denoted  by
$\lame$ and $\lamp$
respectively\footnote{Transverse $e^\mp$ polarizations are irrelevant,
if  the  azimuthal distribution
of the neutralino production plane is integrated over.}.
The contributions to this process at tree level arise from $s$-channel
$Z$-exchange, and $t$- and $u$-channel $\tilde e_L$ and $\tilde
e_R$ exchanges. The  differential cross section may then be
written as \cite{neutralinos1a}
\bqa
\frac{{\rm d} \sigma (e^- e^+ \to \tchi_i \tchi_j)}{{\rm d}
t}& = & \frac{\alpha^2 \pi}{ s^2(1+\delta_{ij})} P_{i11}P_{j11}
 \Big [\Sigma_Z(\lame, \lamp)+
 \Sigma_{\tilde e_L}(\lame, \lamp)
 + \Sigma_{\tilde e_R}(\lame, \lamp)
 \nonumber \\
 &&  +\Sigma_{Z\tilde e_L}(\lame, \lamp)
 +\Sigma_{Z \tilde e_R}(\lame, \lamp)  \Big ] ~,
 \label{chiij-dcross}
\eqa
with the r.h.s. terms representing  respectively  the $Z$-, $\tilde
e_L$- and $\tilde e_R$-square contributions, and  the
$Z\tilde e_L$- and $Z\tilde e_R$-interferences\footnote{To the
extend that we neglect the electron mass, there is never any
$\tilde e_L \tilde e_R$-interference; neither any
Higgsino-$e\tilde e$ coupling.}. These are  written as
\bqa
&& \Sigma_Z(\lame, \lamp) =
\frac{1}{8\sw^4\cw^4 (s-\mzd )^2}
[(g_{ve}^2+g_{ae}^2)(1-\lame\lamp)-2g_{ve}g_{ae}(\lame-\lamp)]
 \nonumber \\
& & \cdot \Bigg \{ |p_{i3}^*p_{j3}-p_{i4}^*p_{j4}|^2
  [(t-m_{\tchi_i}^2)(t-m_{\tchi_j}^2)+
  (u-m_{\tchi_i}^2)(u-m_{\tchi_j}^2)]
  \nonumber \\
 && - 2 s m_{\tchi_i} m_{\tchi_j}
  Re \Big (\eta_i^*\eta_j [p_{i3}^*p_{j3}-p_{i4}^*p_{j4}]^2\Big )
  \Bigg \}  ~ ,
\label{Zsquare}\\
&& \Sigma_{\tilde e_L}(\lame, \lamp)=\frac{1}{4 \sw^4 \cw^4}
\frac{(1-\lame)(1+\lamp)}{4}
\nonumber \\
&& \cdot \Bigg \{ |(\cw p_{i2}+\sw)(\cw p_{j2}+\sw)|^2
    \Bigg  [\frac{(t-m_{\tchi_i}^2)(t-m_{\tchi_j}^2)}
{(t-m_{\tilde e_L}^2)^2} +
\frac{(u-m_{\tchi_i}^2)(u-m_{\tchi_j}^2)} {(u-m_{\tilde e_L}^2)^2}
\Bigg ]
\nonumber \\
&&  - ~\frac{2 s  m_{\tchi_i} m_{\tchi_j}} {(t-m_{\tilde
e_L}^2)(u-m_{\tilde e_L}^2)}
Re [\eta_i^* \eta_j (\cw p_{i2}^*+\sw)^2(\cw p_{j2}+\sw)^2]
\Bigg \}  ~ , \label{seL-square} \\
&& \Sigma_{\tilde e_R}(\lame, \lamp)= \frac{4}{ \cw^4}
\frac{(1+\lame)(1-\lamp)}{4}
 \Bigg \{ \Bigg  [\frac{(t-m_{\tchi_i}^2)(t-m_{\tchi_j}^2)}
{(t-m_{\tilde e_R}^2)^2}
+ \frac{(u-m_{\tchi_i}^2)(u-m_{\tchi_j}^2)} {(u-m_{\tilde e_R}^2)^2}
\Bigg ] \nonumber \\
&& - ~ \frac{2 s  m_{\tchi_i}m_{\tchi_j}}
 {(t-m_{\tilde e_R}^2)(u-m_{\tilde e_R}^2)} Re (\eta_i\eta_j^*)
 \Bigg \} , \label{seR-square} \\
&& \Sigma_{Z \tilde e_L}(\lame, \lamp) = -~ \frac{
(g_{ve}+g_{ae})}{2  \sw^4\cw^4 (s-\mzd)}\frac{(1-\lame)(1+\lamp)}{4}
\Bigg \{ Re \Big  [(p_{i3}^*p_{j3}-p_{i4}^*p_{j4})
\nonumber \\
&& \cdot (\cw p_{i2}+\sw)
(\cw p_{j2}^*+\sw) \Big ]
\Big [\frac{(t-m_{\tchi_i}^2)(t-m_{\tchi_j}^2)}{t-m_{\tilde e_L}^2 }
+\frac{(u-m_{\tchi_i}^2)(u-m_{\tchi_j}^2)}{u- m_{\tilde e_L}^2} \Bigg ]
\nonumber \\
&&  -m_{\tchi_i} m_{\tchi_j} s \Big (\frac{1}{t-m_{\tilde e_L}^2}
+\frac{1}{u-m_{\tilde e_L}^2} \Big )
 Re \Big  [\eta_i \eta_j^*  (p_{i3}p_{j3}^*-p_{i4}p_{j4}^*)
\nonumber \\
&& \cdot  (\cw p_{i2}+\sw) (\cw p_{j2}^*+\sw) \Big ] \Bigg \}
  , \label{ZseL-int} \\
&& \Sigma_{Z \tilde e_R}(\lame, \lamp) =
\frac{ 2(g_{ve}-g_{ae})}{\sw^2\cw^4 (s-\mzd)}
\frac{(1+\lame)(1-\lamp)}{4}
\nonumber \\
&& \cdot \Bigg \{  Re  [p_{i3}p_{j3}^*-p_{i4}p_{j4}^* ]
 \Big [\frac{(t-m_{\tchi_i}^2)(t-m_{\tchi_j}^2)}{t-m_{\tilde e_R}^2}
+\frac{(u-m_{\tchi_i}^2)(u-m_{\tchi_j}^2)}{u- m_{\tilde e_R}^2} \Big ]
\nonumber \\
&& -m_{\tchi_i} m_{\tchi_j} s \Big (\frac{1}{t-m_{\tilde e_R}^2}
+\frac{1}{u-m_{\tilde e_R}^2} \Big )
 Re [\eta_i \eta_j^*  (p_{i3}p_{j3}^*-p_{i4}p_{j4}^*)]
 \Bigg \} ~~  , \label{ZseR-int}
\eqa
where $g_{ve}=-0.5+2\swd$ and $g_{ae}=-0.5$
are the vector and axial $Zee$-couplings.

Close to  threshold, the neutralino-pair production  cross
section  can be expanded
in powers of $\Delta s \equiv s-(m_{\tchi_i}+m_{\tchi_j})^2 $,
as
\bqa
 && \sigma({e^-e^+\to \tchi_i \tchi_j})=
 \frac{\alpha^2 \pi}{ \sw^4 \cw^4 (1+\delta_{ij})}
 \frac{(m_{\tchi_i} m_{\tchi_j})^{3/2}}
 {(m_{\tchi_i}+m_{\tchi_j})^2} P_{i11}P_{j11}
\sqrt{\Delta s}
\nonumber \\
&&  \cdot \Big \{ [(g_{ve}^2+g_{ae}^2)(1-\lame\lamp)
-2g_{ve}g_{ae}(\lame-\lamp)]
\frac{{\cal{K}}_{ij}^2}{((m_{\tchi_i}+m_{\tchi_j})^2-\mzd)^2}
\nonumber \\\ &&+  \frac{(1-\lame)(1+\lamp)}{4} \frac{2
{\cal{J}}_{ij}^2}{(m_{\tchi_i} m_{\tchi_j} +m_{\tilde e_L}^2)^2}+
\frac{(1+\lame)(1-\lamp)}{4} \frac{32 \sw^4
{\cal{I}}_{ij}^2}{(m_{\tchi_i} m_{\tchi_j} +m_{\tilde e_R}^2)^2}
\nonumber \\ && +(g_{ve}+g_{ae}) \frac{(1-\lame)(1+\lamp)}{4}
\frac{4 {\cal{J}}_{ij} {\cal{K}}_{ij}}
{((m_{\tchi_i}+m_{\tchi_j})^2-\mzd)(m_{\tchi_i} m_{\tchi_j}
+m_{\tilde e_L}^2)} \nonumber
\\ &&-(g_{ve}-g_{ae})
\frac{(1+\lame)(1-\lamp)}{4} \frac{16 \swd {\cal{I}}_{ij}
{\cal{K}}_{ij}} {((m_{\tchi_i}+m_{\tchi_j})^2-\mzd)(m_{\tchi_i}
m_{\tchi_j} +m_{\tilde e_R}^2)} \Big \}
\nonumber \\
&&+0 ([\Delta s]^{3/2}) ~~,
\label{chiij-thresh}
\eqa
where
\bqa
{\cal{I}}_{ij}&=& {\rm Im}[\sqrt{\eta_i^* \eta_j}~] ~~, \nonumber \\
 {\cal{J}}_{ij}&=& {\rm Im}[\sqrt{\eta_i^* \eta_j}
 (\cw p_{i2}^*+\sw)(\cw p_{j2}+\sw)] ~~,\nonumber  \\
 {\cal{K}}_{ij}&=& {\rm Im}[\sqrt{\eta_i^* \eta_j}
 (p_{i3}^*p_{j3}-p_{i4}^* p_{j4})]~~ . \label{chiij-thresh-par}
\eqa

Therefore,  the leading behaviour
$\sim \Delta s^{1/2}$,  can only be realized if the  produced
neutralinos are not identical, and such that at least one of
$({\cal I,~ J, ~ K,})$ of (\ref{chiij-thresh-par}),
is non vanishing. The next to leading term varies like
$\Delta s^{3/2}$.

In the special case that CP is conserved
(\ie  $p_{j\alpha}$  real),
the above conditions allow the appearance of the leading
$\Delta s^{1/2}$ term, only if  $\eta_i= -\eta_j=\pm 1$
\cite{Zerwas1}.

But, if instead CP is violated
and the reduced projector elements are complex, then
(\ref{chiij-thresh-par}) could allow the appearance of the
leading threshold term $\Delta s^{1/2}$,
even if  $\eta_i= \eta_j$, provided
of course that the two neutralinos continue to be different.

Finally, we note that the vanishing of the
leading $\Delta s^{1/2}$ term in the
production of two identical neutralinos, may be simply
viewed as a consequence of the fact
that the $(e^-, ~ e^+)$ helicities have to be
opposite to each other. The reason is that the
total angular momentum of the two neutralinos would then  be at least
one, which forces the S-wave neutralino wave
function\footnote{For S-wave, the total
neutralino spin should be one. Neutralino states with
higher  orbital angular momentum cannot produce
a $\Delta s^{1/2}$ term.} to be antisymmetric,  implying
 a threshold behaviour like   $\Delta s^{3/2}$.

\section{Determining SUSY  parameters.}

In this section we address the problem of determining the MSSM
SUSY parameters, under the various conditions that will inevitably
arise when some (presumably the lightest)
  charginos and neutralinos will start
being discovered. In all cases,
we assume that $\tan\beta$
 and some knowledge of the
chargino parameters  $(M_2, ~\bar \mu, ~\Phi_\mu)$,
will have been established  from chargino and other measurements,
before the neutralinos start being studied.
Then, neutralino measurements may supply information
on their physical masses, reduced projector elements
 and  CP eigenphases.

Our general procedure  starts from
(\ref{eqeta2}-\ref{eqeta4}) which are solved for $p_{j2}, p_{j3},
p_{j4} ~ (j=1-4)$, without introducing any explicit
dependence on the Bino parameters $\bar  M_1, \Phi_1$.
This gives
\bqa
\label{z2-CP}
p_{j2}&=&t_W+(m_{\tchi_j}^2-\bar \mu^2) {\cal{Z}}_j^*  ~~, \\
\label{z3-CP}
p_{j3}&=& \mw \ ( \cbeta
\tilde m_{\tchi_j}^* {\cal{Z}}_j
+\sbeta \bar \mu \exb {\cal{Z}}_j^*) ~~, \\
\label{z4-CP}
p_{j4}&=& -\mw  \ (\sbeta \tilde m_{\tchi_j}^*{\cal{Z}}_j
+\cbeta \bar \mu \exb {\cal{Z}}_j^*)
~~ ,
\eqa
where the auxiliary complex numbers ${\cal{Z}}_j~(j=1-4)$
are defined as
\bq
\label{C-aux}
{\cal{Z}}_j={\cal{Z}}_{j1}-\tilde m_{\tchi_j} ~ {\cal{Z}}_{j2} ~~,
\eq
with
\bqa
\label{C-aux1}
 {\cal{Z}}_{j1}&=& t_W {\cal{D}}_j^{-1} [(M_2^2-m_{\tchi_j}^2)
(m_{\tchi_j}^2-\bar \mu^2)+\mwd  (m_{\tchi_j}^2+  M_2
\bar \mu \exb \s2beta)] ~~ , \\
\label{C-aux2}
 {\cal{Z}}_{j2} &=&  \mzd \sw \cw {\cal{D}}_j^{-1} (M_2+\bar \mu
\exb \s2beta) ~~ , \\
\label{D-aux}
 {\cal{D}}_j& =& (m_{\tchi_j}^2-\bar \mu^2)^2
(m_{\tchi_j}^2-M_2^2) \nonumber \\
&& +\mwd [2
(\bar \mu^2-m_{\tchi_j}^2) (m_{\tchi_j}^2+M_2 \bar \mu \s2beta
\cb)+\mwd (m_{\tchi_j}^2-\bar \mu^2 \s2beta^2) ~~.
\eqa
Note that ${\cal{Z}}_{j1},{\cal{Z}}_{j2}, {\cal{D}}_j$ do not depend
on the CP-eigenphase $\eta_j$, but only on the physical neutralino mass
$m_{\tchi_j}=|\tilde m_{\tchi_j}|$.

Substituting (\ref{z2-CP}-\ref{z4-CP}) in (\ref{M1c})
we then obtain
\bq
\label{M1comp}
M_1= \bar  M_1 \exa = A_j \eta_j+B_j ~~,
\eq
where
\bqa
A_j&=&  m_{\tchi_j} {\cal{D}}_j^{-1}[(m_{\tchi_j}^2-M_2^2)
(m_{\tchi_j}^2-\bar \mu^2)^2+\mzd \mwd  (m_{\tchi_j}^2-\bar \mu^2
\s2beta^2) \nonumber \\
&& -\mzd (m_{\tchi_j}^2-\bar \mu^2) (2 m_{\tchi_j}^2
\cwd+(m_{\tchi_j}^2-M_2^2) \swd +2 \bar \mu M_2 \s2beta \cwd \cb)]
~~, \nonumber \\
B_j&=& -{\cal{D}}_j^{-1} \mzd \swd [(m_{\tchi_j}^2-M_2^2)
(m_{\tchi_j}^2-\bar \mu^2)
\bar \mu e^{-i \Phi_\mu} \s2beta+\mwd M_2(m_{\tchi_j}^2-\bar \mu^2
\s2beta^2)  ] ~~, \label{AjBj-CP}
\eqa
and ${\cal{D}}_j$ is given in (\ref{D-aux}).

The set of equations (\ref{z2-CP}-\ref{AjBj-CP}) determines
$M_1$ including its phase, under various conditions
concerning the knowledge of the neutralino
and chargino sectors.

\vspace{0.2cm}
If  CP  is conserved; then
 $p_{j\alpha}$, $\eta_j$ and the SUSY parameters are
 all real, so
that  (\ref{z2-CP}-\ref{z4-CP}, \ref{M1comp}) become
\bqa
p_{j2}&=& \frac{ \mz \mw (\tilde
m_{\tchi_j}+\s2beta  \mu) \sw }{(M_2-\tilde m_{\tchi_j})
(\tilde m_{\tchi_j}^2-\mu^2) +\mwd  (\tilde
m_{\tchi_j}+\s2beta  \mu)}~~, \label{z2-CPcon} \\
p_{j3}&=& -\frac{ \mz (M_2-\tilde m_{\tchi_j}) (\cbeta \tilde
m_{\tchi_j}+\sbeta
\mu) \sw}{(M_2-\tilde m_{\tchi_j}) (\tilde m_{\tchi_j}^2-\mu^2)
+\mwd  (\tilde m_{\tchi_j}+\s2beta  \mu)}
~~, \label{z3-CPcon} \\
p_{j4}&=&\frac{ \mz (M_2-\tilde m_{\tchi_j}) (\sbeta \tilde
m_{\tchi_j}+\cbeta  \mu) \sw}{(M_2-\tilde m_{\tchi_j})
(\tilde m_{\tchi_j}^2-\mu^2) +\mwd  (\tilde
m_{\tchi_j}+\s2beta  \mu)} ~~, \label{z4-CPcon}
\eqa
\bqa
M_1 &=&
\tilde m_{\tchi_j}+\mz \sw (p_{j3} \cbeta-p_{j4} \sbeta)
\nonumber  \\
&=& \frac{\tilde m_{\tchi_j} (\tilde m_{\tchi_j}-M_2) (\tilde
m_{\tchi_j}^2-\mu^2)-\mzd ( m_{\tchi_j} +\s2beta \mu)
(\tilde m_{\tchi_j}-M_2 \swd)}{(\tilde m_{\tchi_j}^2-\mu^2)
(\tilde m_{\tchi_j}-M_2)-\mwd (\tilde m_{\tchi_j}+\s2beta
\mu)} ~~, \label{M1-CPcon}
\eqa
where  the "signed" neutralino
 masses defined in  (\ref{complex-masses}) (of course for real $\eta_i$)
  appear.

\vspace{0.2cm}
 We next consider various scenarios in the CP conserving and CP
 violating cases for the neutralino-chargino sectors.

\subsection{CP-conserving scenarios}

\noindent
$\bullet$ The scenario {\cal S}1. \\
This is a rather extreme scenario considered in \cite{Kneur-CPc},
 where we assume    that the real MSSM
parameters $(M_2, ~\tan \beta, ~\mu)$ are known from \eg
the chargino sector. In addition to them,   the
signed mass $\tilde m_{\tchi_i}=\eta_i m_{\tchi_i}$
of one neutralino  $\tchi_i$ is also assumed known.
Knowledge of the value of the index $i$ determining the ordering
of $\tchi_i$, is not assumed. Then,
(\ref{M1-CPcon}) allows us to
reconstruct  $M_1$, including its sign.

\begin{figure}[t]
\vspace*{-3cm}
\[
\epsfig{file=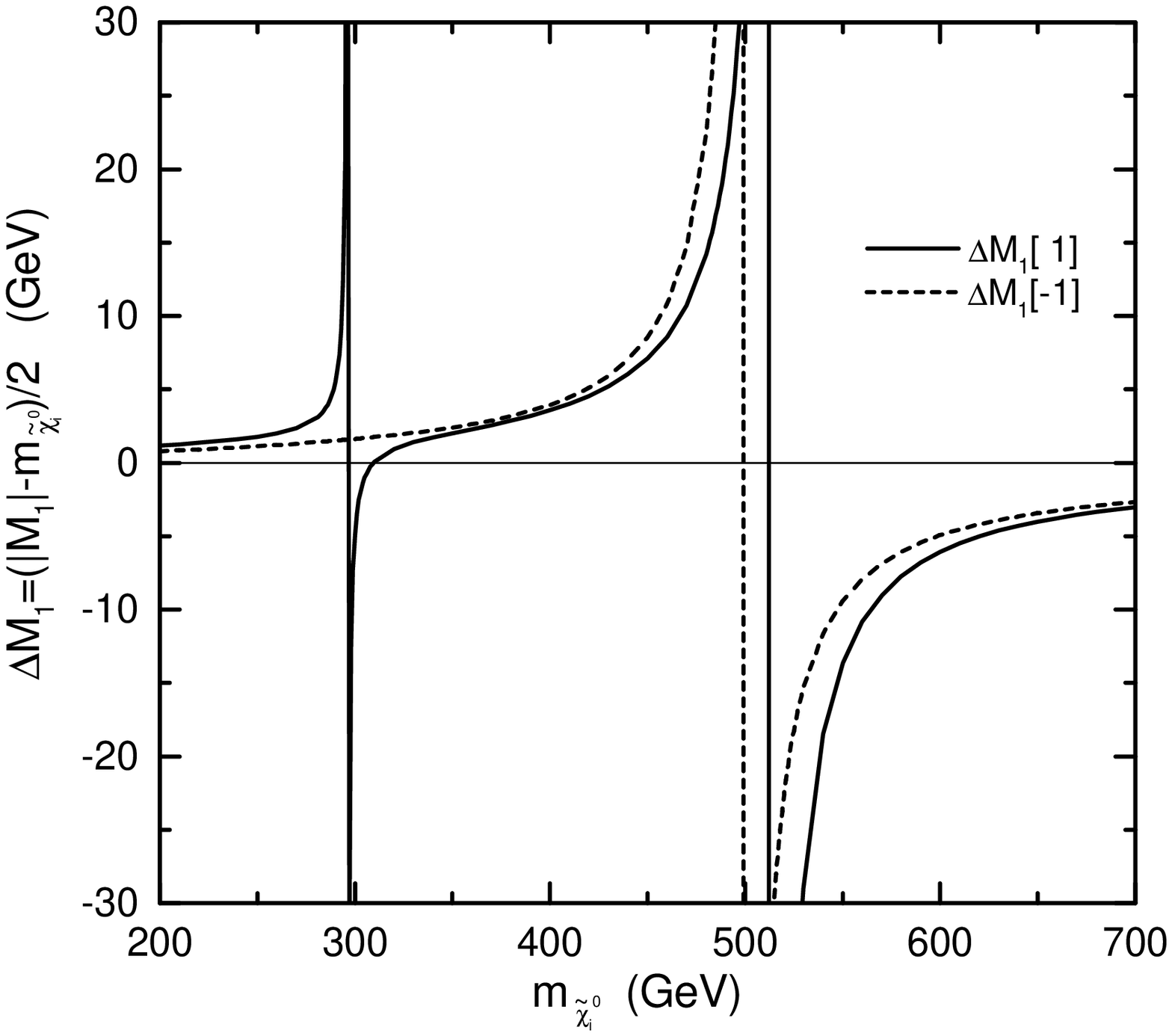,height=9.cm}
\]
\vspace*{-1.cm}
\caption[1]{Reconstruction of $M_1$ in the  {\cal S}1
 CP-conserving scenario in which  $M_2$, $\mu$,
$\tan \beta$ are taken as in model SP1b of Table 1.
 $|M_1|$ is determined from (\ref{M1-CPcon}), for
 varying $m_{\tchi_i}$. Solid (broken) lines correspond
to $\eta_i=1~ (-1)$ respectively.
For lower  $m_{\tchi_i}$-values than those indicated, the
results remain the same as for $m_{\tchi_i}\sim 200$ ~GeV. }
\label{M1-CPcon-fig}
\end{figure}

It may turn out though, that only the physical mass
of $\tchi_i$  is known, and not its CP eigensign  $\eta_i$.
As an example,
we consider  the model $SP1b$ of Table 1, which is
essentially identical to the Snowmass benchmark SPS1b
\cite{Snowmass}.
In this model,  one neutralino is
almost  a pure Bino of mass $|M_1|$, another one
  is approximately a Wino of mass $M_2$, and the last
  two are almost degenerate  Higgsinos
of mass  $\sim |\mu |$. This property seems rather stable
under model variations where ($M_2$, $\tan\beta$,  $ \mu$)
are kept fixed, while the eigensign $\eta_i$ is changed.
To see this, we present  in Fig.\ref{M1-CPcon-fig}  the values of
$(|M_1|-m_{\tchi_i})/2$ predicted by (\ref{M1-CPcon})
as a function of  $m_{\tchi_i}$, for the two choices
$\eta_i=\pm 1$. It seems from this
figure that if $(M_2, ~\mu)$ are known, and the
mass of the considered  neutralino  is not very close to either
$M_2$ or $|\mu|$; then it should  be almost identical to $|M_1|$,
for any $\eta_i$ sign.

\begin{figure}[htb]
\vspace*{-3.0cm}
\[
\hspace{-0.5cm}
\epsfig{file=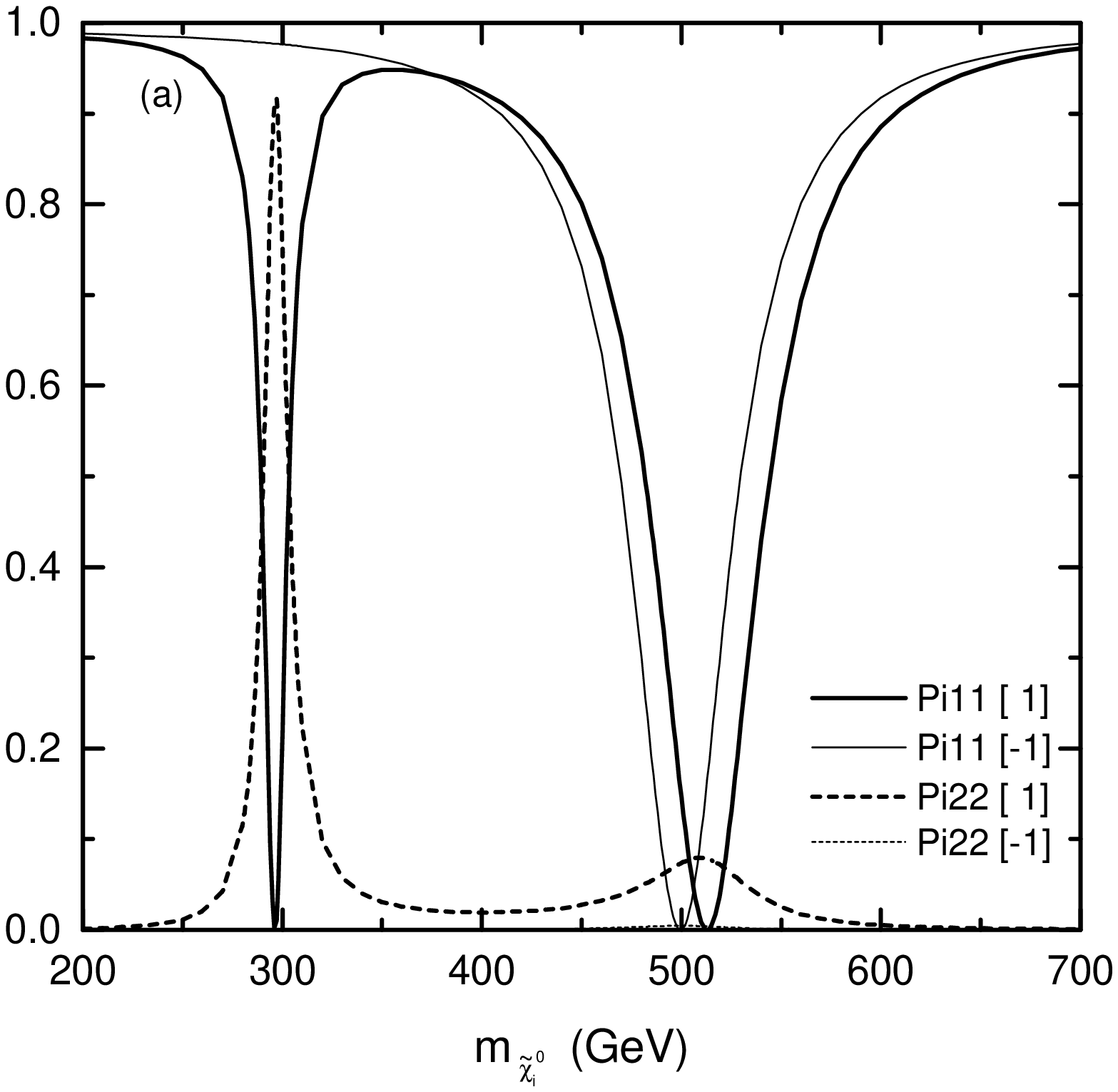,height=8.5cm}\hspace{-1.0cm}
\epsfig{file=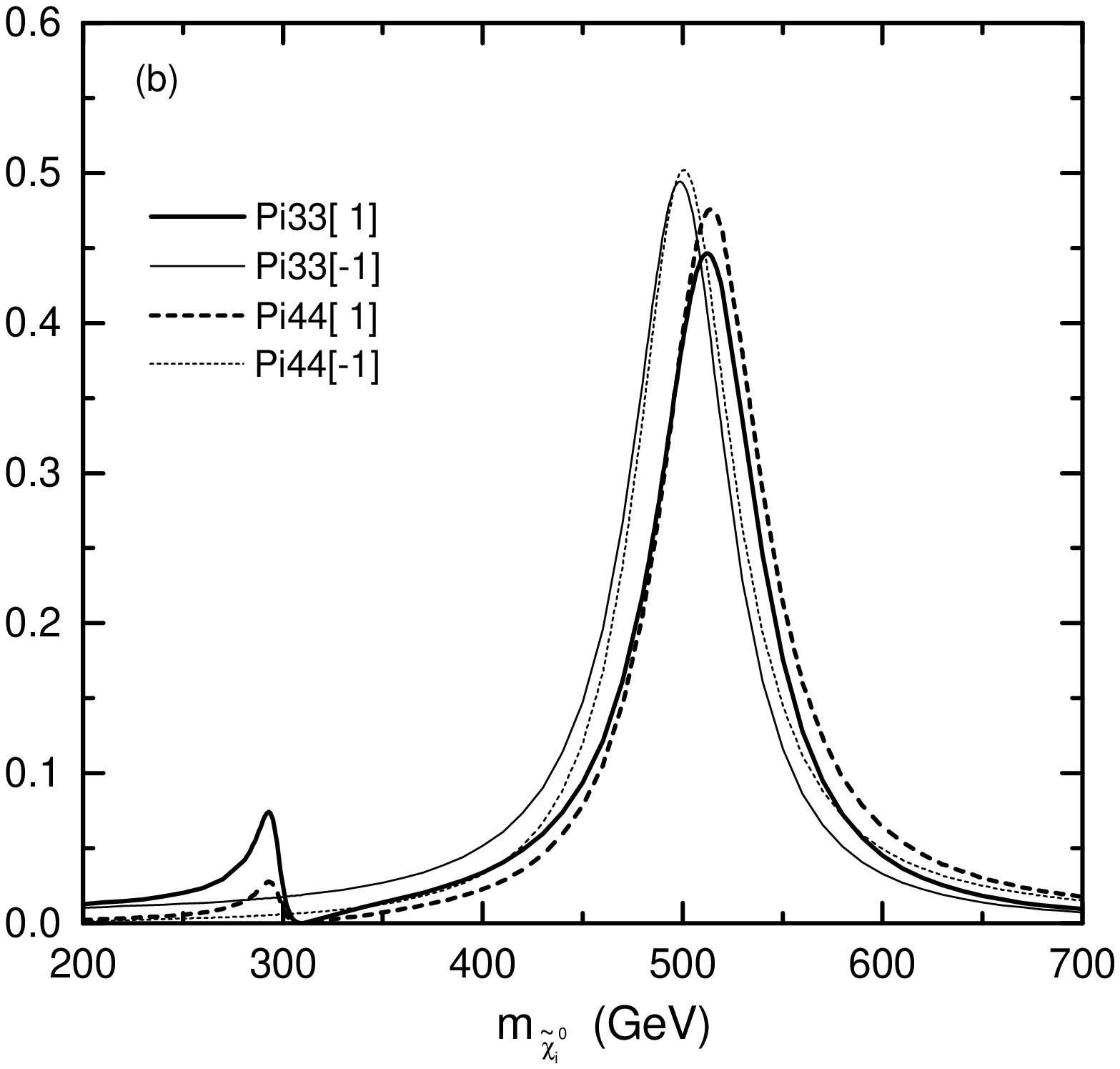,height=8.5cm}
\]
\vspace*{-1.0cm}
\caption[1]{Diagonal projector elements $P_{i11}$,
$P_{i22}$ (a) and $P_{i33}$, $P_{i44}$ (b), as  a function of
the physical mass $m_{\tchi_i}$,  for the same parameters as in
Fig.\ref{M1-CPcon-fig}. The values of $P_{i22}[-1]$ are so small
that they cannot be seen.  For lower  $m_{\tchi_i}$-values
than those indicated, the
results remain the same as for $m_{\tchi_i}\sim 200$ ~GeV.}
\label{proj-fig}
\end{figure}

For the same data,  we show in Fig.\ref{proj-fig}, the
diagonal elements of $P_i$ as a function of $m_{\tchi_i}$, for the
two possible values $\eta_i=\pm 1$. This figure confirms
that if  $m_{\tchi_i}$ is close to $|M_1|$,
then   $\tchi_i$
is approximately  a pure Bino state.

The lack of knowledge of $\eta_i$ does not seem important,
at least in the above example. Nevertheless,  the
ambiguity induced by it may  be lifted, provided some
additional information is used:

One possibility could be that a second neutralino with mass
 $m_{\tchi_j} ~(j\neq i)$ is also known.
The  ambiguity is then lifted by comparing the values of $M_1$
obtained when  trying all possible choices
for $\eta_i=\pm 1$ and $\eta_j=\pm 1$. The correct
$(\eta_i, ~ \eta_j)$ choice, is the one
which leads to the \underline{same} $M_1$ prediction.

 Another possibility arises from the fact that typically
 the  $M_1$ values, obtained for $\eta_i=\pm 1$, are
 of opposite signs. Therefore, if a
particular sign for $M_1$ is favored (as \eg in mSUGRA models
with gaugino unification for which
${\rm Sign}[M_1]={\rm Sign}[M_2]$ is expected), the ambiguity is
again  lifted.

A third possibility of determining $\eta_j$, is from
$\sigma(e^- e^+\to \tchi_j \tchi_i)$ for $j \neq i$,
or from chargino decays
to neutralinos, if such measurements exist.

\vspace{0.3cm}
\noindent
$\bullet$ The scenario {\cal S}2. \\
In this  scenario, it is assumed that the first quantities to
be measured  will be the physical masses of one
 chargino and two neutralinos  \cite{Kneur-CPc}. As such, we take  the
 mass of the lightest chargino  $\tchic_1^\pm $, and
 the two lightest neutralino masses. If the signed neutralino masses
$\tilde m_{\tchi_1}$ and
$\tilde m_{\tchi_2} $
( compare (\ref{complex-masses})),  were
also  known\footnote{We come back to the determination
of the signs $\eta_1$ and $\eta_2$   at the end
of this subsection.}, then    (\ref{M1-CPcon})
  would give the $M_2$-quadratic equation
\bq
\label{M2-neut}
a_M^{(0)} + a_M^{(1)} M_2+a_M^{(2)} M_2^2=0 ~~,
\eq
whose coefficients
\bqa
&& a_M^{(0)} = \tilde m_{\tchi_1} \tilde
m_{\tchi_2} (\tilde m_{\tchi_1}-\tilde m_{\tchi_2}) (\tilde
m_{\tchi_1}^2-\mu^2) (\tilde m_{\tchi_2}^2-\mu^2)
+ \mzd [\tilde m_{\tchi_1}\tilde m_{\tchi_2} \{ (\tilde
m_{\tchi_1}-\tilde m_{\tchi_2}) (\tilde m_{\tchi_1} \tilde
m_{\tchi_2} \nonumber \\
&& + (1+\cwd) \mu^2+\mzd \cwd)+\cwd (\tilde
m_{\tchi_2}^3-\tilde m_{\tchi_1}^3)\}
 + \mu \ \s2beta (\tilde m_{\tchi_1}-\tilde m_{\tchi_2})
 \{ (\tilde m_{\tchi_1}+\tilde m_{\tchi_2}) (\tilde m_{\tchi_1} \tilde
m_{\tchi_2} \nonumber \\
&& -\cwd (\tilde m_{\tchi_1}^2+\tilde
m_{\tchi_2}^2-\mu^2-\mzd)) + \mu \mzd \s2beta \cwd \}]
~~~,  \nonumber \\
&& a_M^{(1)} = (\tilde m_{\tchi_2}^2-\tilde m_{\tchi_1}^2)
 (\tilde m_{\tchi_1}^2-\mu^2)
(\tilde m_{\tchi_2}^2-\mu^2) + \mzd [(\tilde
m_{\tchi_1}^2-\tilde m_{\tchi_2}^2) ((\cwd-\swd) \tilde
m_{\tchi_1} \tilde m_{\tchi_2}-\mu^2) \nonumber \\
&& +\mu \s2beta ((\tilde m_{\tchi_1}-\tilde m_{\tchi_2}) (\tilde
m_{\tchi_1}^2+\tilde m_{\tchi_2}^2-2 \mu^2  \cwd)+2 (\tilde
m_{\tchi_2}^3-\tilde m_{\tchi_1}^3) \swd)]~~ , \nonumber \\
&& a_M^{(2)} =  (\tilde m_{\tchi_1}-\tilde m_{\tchi_2}) (\tilde
m_{\tchi_1}^2-\mu^2) (\tilde m_{\tchi_2}^2-\mu^2) \nonumber \\
&& +\mzd \swd [\tilde m_{\tchi_2} (\tilde m_{\tchi_1}^2-\mu^2)
-\tilde m_{\tchi_1} (\tilde m_{\tchi_2}^2-\mu^2)+\mu \ \s2beta
(\tilde m_{\tchi_1}^2-\tilde m_{\tchi_2}^2)] ~~,
\label{M2-neut-coef}
\eqa
depend on $\mu$ in a complicated way.

A second  independent quadratic $M_2$-equation, with
coefficients depending on $\tan \beta$, $\mu$ and
\eg the lightest chargino mass\footnote{In our convention,
 the physical (positive) chargino masses are ordered so that
$m_{\tchic_1^\pm} < m_{\tchic_2^\pm}$.}  $m_{\tchic_1^\pm}$,   may be
derived by noting that the chargino mass matrix in the CP
conserving case implies
\bqa
&& m_{\tchic_1^\pm}^2+m_{\tchic_2^\pm}^2=M_2^2+\mu^2+2 \mwd
 ~~ , \nonumber \\
&& m_{\tchic_1^\pm}^2 m_{\tchic_2^\pm}^2=(M_2 \mu- \mwd \s2beta)^2
~~ . \label{chargino-masses}
\eqa
Eliminating  the heavier chargino mass then leads to
\bq
\label{M2-charg}
M_2^2 (m_{\tchic_1^\pm}^2-\mu^2)+2 \mu \mwd \s2beta
M_2-m_{\tchic_1^\pm}^4+m_{\tchic_1^\pm}^2 (\mu^2+2 \mwd)-\mw^4
\s2beta^2=0 ~~ .
\eq
Combining (\ref{M2-neut}, \ref{M2-neut-coef}, \ref{M2-charg}),
we   find
\bq
\label{M2-comp}
M_2=-\frac{a_M^{(0)} (m_{\tchic_1^\pm}^2-\mu^2)+a_M^{(2)}
(m_{\tchic_1^\pm}^4-m_{\tchic_1^\pm}^2 (\mu^2+2 \mwd)+\mw^4
\s2beta^2)}{a_M^{(1)} (m_{\tchic_1^\pm}^2-\mu^2)-2 a_M^{(2)} \mu
\mwd \s2beta} ~,
\eq
and the compatibility relation
\bqa
\label{mu-comp}
&&[a_M^{(2)} (m_{\tchic_1^\pm}^4-m_{\tchic_1^\pm}^2
(\mu^2+2 \mwd)+\mw^4
\s2beta^2)+a_M^{(0)} (m_{\tchic_1^\pm}^2-\mu^2)]^2-
[a_M^{(1)} (m_{\tchic_1^\pm}^2-\mu^2) \nonumber \\
&& -2 a_M^{(2)} \mwd \mu \s2beta] [a_M^{(1)} (m_{\tchic_1^\pm}^4-
m_{\tchic_1^\pm}^2 (\mu^2+2 \mwd)+\mw^4 \s2beta^2)+2 a_M^{(0)}
\mwd \mu \s2beta]=0 ~~.
\eqa

Our next task is to  solve (\ref{mu-comp}) for $\mu$,
in terms of  $\tan \beta$, the lightest physical chargino mass
$m_{\tchic_1^\pm}$, and the signed neutralino masses $\tilde
m_{\tchi_1}$ and $\tilde m_{\tchi_2}$.
In principle there could be many solutions, since
(\ref{mu-comp}) is of order 12 in $\mu$.
Their number is reduced though by restricting to
 real $\mu$,  such that (\ref{M2-comp}) leads to
positive $M_2$. Further restrictions arise
by calculating also the masses of the other charginos and
neutralinos, and  demanding that those
we have started with,  are indeed the lightest ones.

As an example we consider again the benchmark mSUGRA
 models $SP1b$ of \cite{Snowmass} and\footnote{The interest in
 this model is that it features a negative $\mu$.}
 $D$ of \cite{Ellis-bench}, whose
 relevant electroweak scale parameters, calculated using the
 SUSPECT code\footnote{Of course the conclusions below
 do not depend on the specific use of this code.}
 \cite{SUSPECT},  are presented in Table 1.
 We intend to look at how these models are reconstructed (always of
 course at the   electroweak scale), using
  in the context of scenario {\cal S}2,  the formalism
  developed above.
Solving (\ref{mu-comp}) for $\mu$ in the  $SP1b$ ~($D$) case,
we find 8 ~(10) real solutions, out of which only
 6 ~(7) are  consistent with of our starting convention $M_2>0$.
 Finally, when imposing the additional
 requirement that the solutions   respect
 the starting   mass hierarchies;
 then only two remain for each of the $SP1b$ and $D$ cases.

 Thus, apart from the models
 $SP1b$ and $D$ which are of course reconstructed, two new
 solutions are found denoted as  $SP1b'$ and $D'$
 in Table 1. By construction therefore, the pairs of models
($SP1b$,  ~$SP1b'$) and ($D$,  ~$D'$),
have the same mass for $\tchic^\pm_1$, and the
same signed masses for $\tchi_1, ~ \tchi_2$, as well as the same
value for $\tan\beta$.
But their derived   $M_1, ~M_2, ~\mu$  parameters and
the masses for $\tchic_2^+$, $\tchi_3$, $\tchi_4$,
 are generally  different.

\begin{table}[h]
\begin{center}
{ Table 1: Electroweak scale parameters for some
CP conserving models; (Masses  in $\rm GeV$.)}\\
  \vspace*{0.3cm}
\begin{tabular}{||c||c|c|c|c||c|c|c|c|c|c||}
\hline \hline
Model & $\tan \beta$ & $\mu$ & $M_1$ & $M_2$ &
$m_{\tchic_1^\pm}$ & $m_{\tchic_2^\pm}$ & $\tilde m_{\tchi_1}$ &
$\tilde m_{\tchi_2}$ & $\tilde m_{\tchi_3}$ &  $\tilde
m_{\tchi_4}$
\\ \hline
$SP1b$ & 30 &  496 & 166 & 310 &
297& 516 & 164 & 297 & -501 & 516 \\
$SP1b'$ & 30 & -398 & 166 & 320 &297
& 431 &164 &297 & -405 & 430 \\
$SP1b''$ & 30 & 298  & -168 & 1156 & 297
& 1162 &- 164 &297 & -307 & 1162 \\
$SP1b'''$ & 30 & -297  & -170 & 1627 & 297
& 1631 &- 164 &297 & -306 & 1631
 \\ \hline
$D$ & 10 &  -659 & 225 & 418 & 411 & 673 & 224 &
411 & -664 & 671
\\
$D'$ &10 & -430 & 226 & 519 &411 & 546 &224  &411  & -436 & 546  \\
$D''$ &10 & -765 & -226 & 415 &411 & 776 &-224 &411 & -771 & 773  \\
$D'''$ &10 & 639 & -225 & 425 & 411 & 658& -224 &411 & -644 & 657  \\
$D^{IV}$ &10 & 416 & -226 & 973 & 411 & 982& -224 &411 & -422 & 982 \\
$D^{V}$ &10 & -411 & -229 & 2673 & 411 & 2676 & -224 &411 & -418 & 2676 \\
\hline \hline
\end{tabular}
\end{center}
\end{table}

 We next turn to the CP eigensigns  $\eta_1=\pm 1$
 and $\eta_2=\pm 1$,   which  are needed in order to determine
 the signed masses from the physical ones.
The lack of knowledge of these
 signs creates additional ambiguities whose characteristics should
 be studied on a-case-by-case basis.
The aforementioned models   $SP1b$, $SP1b'$, $D$ and
$D'$ in Table 1,  correspond to the choice $\eta_1=\eta_2=+1$.

To study the consequences of the eigensign choice,
we consider  in Table 1 variations of
 the $SP1b$ and $D$ models  in which the same
 \underline{physical} masses
  for the lightest chargino and the two lightest
 neutralino states are used,  but
 we have set instead $\eta_1=-\eta_2=-1$.
We then find   two new solutions  for $(M_1, ~ M_2, ~\mu)$
 and the heavier chargino and neutralino masses,
  denoted as $SP1b''$ and $SP1b'''$ and  displayed in Table 1.
  The same procedure for the $D$-case, produced four
new   solutions, denoted as $D''$, $D'''$,
$D^{IV}$, $D^{V}$ and   shown in Table 1.  We note that in the models
 $SP1b''$, $SP1b'''$, $D^{IV}$ and $D^{V}$, the second and third
 neutralinos are almost degenerate and considerably smaller
than the heaviest  one. Another feature is that  the
two chargino masses tend  to be very different.

  In the same spirit, we have also tried
  variations  to the   $SP1b$ and $D$ models with
  $\eta_1=-\eta_2=1$, but   no consistent solution
  was found. It seems therefore that for an initial  choice of the
physical  masses, not all $\eta_j$-signs are generally possible.

\begin{figure}[p]
\vspace*{-3cm}
\[
\epsfig{file=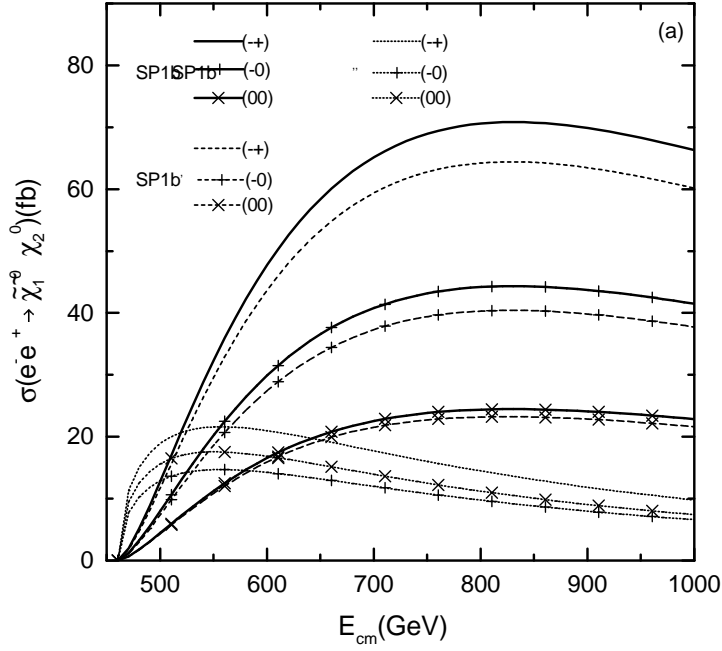,height=11.cm}
\]
\vspace*{-0.0cm}
\[
\epsfig{file=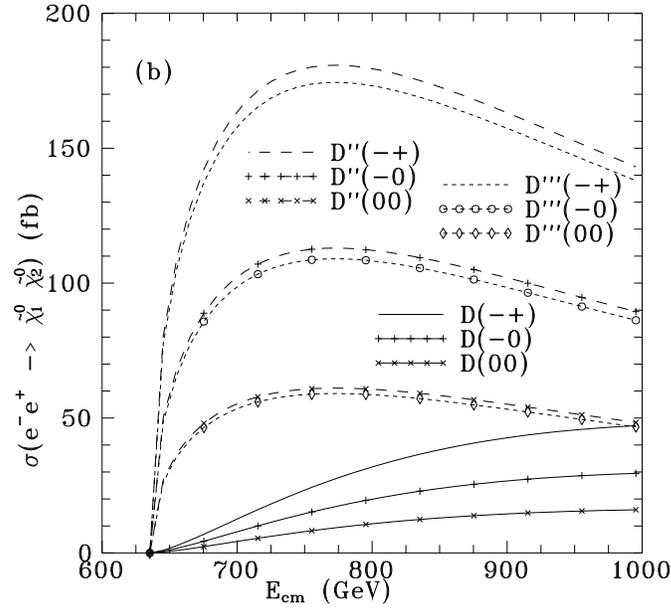,height=8.cm}
\]
\vspace*{-0.5cm}
\caption[1]{Total $\sigma(e^-e^+\to \tchi_1 \tchi_2)$ cross sections
for the CP conserving models $SP1b$, $SP1b'$,
and $SP1b''$  (a), and $D$,  $D''$, $D'''$ (b); see Table 1.
The beams are taken either  unpolarized, or with longitudinal
polarizations of $\lambda_1=\pm 0.85$ for $e^-$  and
$\lambda_2=\pm 0.6$ for $e^+$, indicated
by ${\rm Sign}(\lambda_j)$. The not plotted cross sections
for the polarizations (+0) and (+-) are considerably smaller.
For clarity, we also note that the marks on the (-0) lines,
are actually pluses (+).}
\label{sigma-CPc-fig}
\end{figure}

An efficient way of constraining    $\eta_1$, $\eta_2$, and
the other model parameters, is by looking at the
$\sigma(e^- e^+\to \tchi_1 \tchi_2)$ cross section,
preferably using polarized
beams. In Fig.\ref{sigma-CPc-fig}a and \ref{sigma-CPc-fig}b
we present these
cross sections for the CP-conserving
models ($SP1b$, $SP1b'$, $SP1b''$) and
($D$, $D''$, $D'''$) respectively. The necessary input data
are given in Table 1, while the reduced projector elements are
calculated either from (\ref{z2-CPcon}-\ref{z4-CPcon}) or
(\ref{pj-variables1-CPc}). For the $\tilde e_L, ~\tilde e_R$
masses we used  $m_{\tilde e_L}=338~GeV$, $m_{\tilde e_R}=251.3~GeV$
for the  $SP1b$-type models \cite{Snowmass},
and $m_{\tilde e_L}=385.8~GeV$, $m_{\tilde e_R}=240.2 ~GeV$
for the  $D$-type ones \cite{Ellis-bench}.

 As seen in Figs. \ref{sigma-CPc-fig}a, b,
there is a striking difference
between the cross sections for the models ($SP1b$, $SP1b'$, $D$)
which have $\eta_1 \eta_2=1$, and the models ($SP1b''$, $D''$,
$D'''$) which have $\eta_1 \eta_2=-1$.
Close to the $\tchi_1 \tchi_2$ production threshold, such a
difference may be understood on the basis of
(\ref{chiij-thresh}, \ref{chiij-thresh-par}), which suggest that
the bigger cross section arises for models with
$\eta_i \eta_j=-1$. At higher energies, the
relative magnitude of the various polarization cross sections
is model dependent. The threshold  behaviour  is preserved
as  the  energy  increases,  in
Fig.\ref{sigma-CPc-fig}b, but not  in
Fig.\ref{sigma-CPc-fig}a.

Since the neutralinos are rather heavy
 in the benchmark models \cite{Snowmass, Ellis-bench} we
consider here, we contemplate a Linear
Collider at 800 GeV c.m. energy, with an integrated
luminosity of about $500~ fb^{-1}$ \cite{TESLA3}.
It should then be  possible to discriminate models with
opposite values of $\eta_1 \eta_2$; compare $SP1b$ or $SP1b'$ with
$SP1b''$ in Fig.\ref{sigma-CPc-fig}a; or $D$ with $D''$ or $D'''$
in Fig.\ref{sigma-CPc-fig}b\footnote{Similar remarks should apply
also to models $SP1b'''$, $D^{IV}$, $D^{V}$.}.

Even the discrimination between models with the same value
of $\eta_1 \eta_2$ may be possible,
at least if some knowledge of the
$\tilde e_L$, $\tilde e_R$ masses is
available\footnote{Ideas on how this could be achieved are
presented in \cite{Bartl3, Zerwas1}}.
This is inferred from the fact that for a $500fb^{-1}$ luminosity,
the difference between the
 unpolarized cross sections at 800 GeV for the models $SP1b$
and $SP1b'$ is at the level of 6.7 statistical Standard
Deviations (SD); and
it increases to 12 SD when longitudinally polarized beams are used
with  $\lambda_1= -0.85$, and  $\lambda_2=+0.6$;
see Fig.\ref{sigma-CPc-fig}a. The corresponding difference between the unpolarized cross
sections  for the $D''$ and $D'''$ models,
(which both have $\eta_1 \eta_2=-1$),   is at the level of 5.7
 statistical SD, and it increases to about 11 SD
for  longitudinal polarizations like those used
in Fig.\ref{sigma-CPc-fig}b.
Of course such signals
will be reduced considerably, when  the additional uncertainties due to
the slepton masses and the experimental systematics are taken into account.

The discrimination between the models $SP1b''$
and $SP1b'''$, or $D^{IV}$ and $D^{V}$,
which could lift the remaining reconstruction ambiguities,
 may  be possible if an accurate determination
of the third neutralino mass is available.
This  may be obtained by measurements
of  the $\sigma(e^- e^+\to \tchi_1 \tchi_3)$
cross section using polarized beams.

Finally, we note from Fig.\ref{sigma-CPc-fig}a, that not only
the overall magnitude, but also the
relative magnitudes of the various polarized  cross sections may
change, as  $\eta_1 \eta_2$ changes.
Therefore,   the measurement of the
physical masses of the lightest chargino and  the two lightest
neutralinos, combined with some knowledge of the
$\sigma(e^- e^+\to \tchi_1 \tchi_2)$  cross section
(and possibly also the measurements of $m_{\tchi_3}$
and  $\sigma(e^- e^+\to \tchi_1 \tchi_3)$),  may
be able to provide more or less unique results for
$M_2, ~ M_1, ~\mu$, \cite{Zerwas1, Kneur-CPc}.
At least this is what happens in the CP conserving
case considered above.

 A nice feature of the present approach
 is that   (\ref{mu-comp}) and  (\ref{M2-comp}),
offer an  analytical way of  disentangling  the
  $\mu$ and  $M_2$ parameters.

\subsection{The CP violating case}

We now address   the CP-violating case in which  there exist
 six parameters affecting the ino sector; namely the set
($M_2$, $\tan \beta$, ~$\bar \mu$, $\Phi_\mu$) affecting both charginos
and neutralinos,  and the set ($\bar M_1$, $\Phi_1$) which
 influences only the neutralinos.

As shown in \cite{Zerwas1, Kneur-CPv}, if the charginos
are not too heavy,  a detailed measurement of
$\sigma(e^- e^+ \to \tchic^+_i \tchic^-_j)$ at a Linear
Collider  with polarized beams,  should be able to determine
the two chargino masses $m_{\tchic_1}^\pm, ~m_{\tchic_2}^\pm$,
as well as the mixing angles $\phi_L, ~\phi_R$ entering the
 diagonalization of the chargino mass matrix. If this is achieved,
 then ($M_2$, $\tan \beta$, ~$\bar \mu$, $\Phi_\mu$) are fully
 determined \cite{Kneur-CPv, Abdel-CPv1}. The measurement of
 neutralino production cross sections could then be used to measure the
 genuine $U(1)$ parameters $|M_1|$ and $\Phi_1$.

 Of course it may turn out that such a full separation of the
 information sources for the two aforementioned sets,
   is not  possible.
 It may for example happen that $m_{\tchic^+_2}$ is too heavy to
 be produced on its mass shell in a Linear Collider.
 We therefore  explore various  scenarios
 concerning the chargino and neutralino measurements:

\vspace{0.3cm}
\noindent
$\bullet$ The scenario CPv1:\\
This is analogous to the {\cal S}1 scenario
of the  CP conserving case \cite{Kneur-CPc}.
We assume in it  that the parameters
($ M_2$, $\tan \beta $, $\bar \mu$, $\Phi_\mu $)
will be measured first from \eg the chargino sector.
Concerning the neutralinos,
the assumption is that  initially  we will only know
the physical mass $m_{\tchi_j}$ of \underline{one}
(not necessarily the lightest) neutralino,
and may also have a  limited knowledge of its
 associated CP-eigenphase $\eta_j $.
The problem  will  then be how to   determine
$|M_1|$ and at least constrain the phase angle $\Phi_1$
from (\ref{M1comp}).

 A feeling of the possibilities concerning  these phases
 may be obtained by looking
at CP-violating variations of  \eg the model $SP1b$, in which the
values of $\Phi_1, ~\Phi_\mu$ are varied as
in\footnote{The same model $SP1b$ appears both in Table 1
and in Table 2; but in the latter Table more
 significant digits are kept in order to indicate the small
 mass changes caused by the changes in the phase angles. } Table 2,
 while  $|M_1|$, $M_2$, $|\mu|$ and $\tan \beta$ are kept fixed.
The models thus obtained are labeled as $SP1c$, $SP1d$,
$SP1e$, $SP1g$. In the same Table 2 we also give
the implied physical masses and $\eta_j$ phases for the two
lightest neutralinos calculated  from the formulae in Appendix
A, and (\ref{pj-variables1}, \ref{eqeta1}).
There are in fact two
 models labeled as $SP1c$ (and another two as  $SP1d$),
 corresponding to the upper and
 lower signs for  $\Phi_1$, ${\rm Arg}\eta_1$ and
 ${\rm Arg}\eta_2$.
 In all models listed  in Table 2,
the dimensional quantities
are  essentially the same, but the phases change considerably.

\begin{table}[h]
\begin{center}
{Table 2: Parameters for some  CP-violating variations of
$SP1b$ of Table 1. \\
(Phase angles  in ${\rm rad}$ and masses in $GeV$,
always at the electroweak scale.)}\\
  \vspace*{0.3cm}
\begin{small}
\begin{tabular}{||c|c|c|c|c|c|c|c||}
\hline \hline
\multicolumn{8}{||c||}
{ $\tan \beta=30$,~~ $ |\mu|=495.9 $, ~~ $M_2=309.6 $}\\
\multicolumn{8}{||c||}
{$m_{\tilde e_L}=338$~~ , ~~ $m_{\tilde e_R}=251.3$} \\
 \hline
Model  & $|M_1|$ & $\Phi_1$ & $\Phi_\mu$ & $m_{\tchi_1}$ &
$ m_{\tchi_2}$ & ${\rm Arg}[\eta_1]$ & ${\rm Arg}[\eta_2]$
 \\ \hline
$SP1b$ & 165.7 & 0 & 0 & 163.9 & 296.7  & 0 & 0 \\
$SP1c$ & 165.7& $\pm \frac{\pi}{3}$ & 0 & 164.1& 296.6
& $\pm 1.049 $ & $\pm 0.718$  \\
$SP1d$ & 165.7 & $\pm \frac{\pi}{3}$ & $\pi$ & 164.4 & 299
& $\pm 1.046$ & $\pm 0.712$ \\ \hline
$SP1e$ & 165.7& $ \frac{\pi}{3}$ & $ \frac{\pi}{4}$ & 164.3
& 296.9 & 1.049 & 0.662  \\
$SP1f$ &166 & -1.33  & $ \frac{\pi}{4}$ & 164.3
& 296.9  & -1.33 & -0.947 \\ \hline
$SP1g$ &165.7&  $ \frac{\pi}{3}$ & $-\frac{\pi}{4}$ & 163.9
& 296.9 & 1.0478 & 0.768  \\
$SP1h$ &165.4 & -0.761 & $-\frac{\pi}{4}$ & 163.9
& 296.9  & -0.763 & -0.483  \\ \hline  \hline
\end{tabular}
\end{small}
\end{center}
\end{table}

In the context of the CPv1 scenario, we concentrate on
the models $SP1b$,  $SP1c$, $SP1d$, $SP1e$, $SP1g$ of Table 2.
We now invert the logic within each of these models, and
 assume that only
$(M_2, ~\tan \beta, ~\bar \mu, ~\Phi_\mu)$ and the physical mass
of  one neutralino  $\tchi_j$, are known. Under these conditions, we
explore the extent to which $|M_1|$ and $\Phi_1$ may be
reconstructed.

\begin{figure}[t]
\vspace*{-2cm}
\[
\hspace{-0.5cm}
\epsfig{file=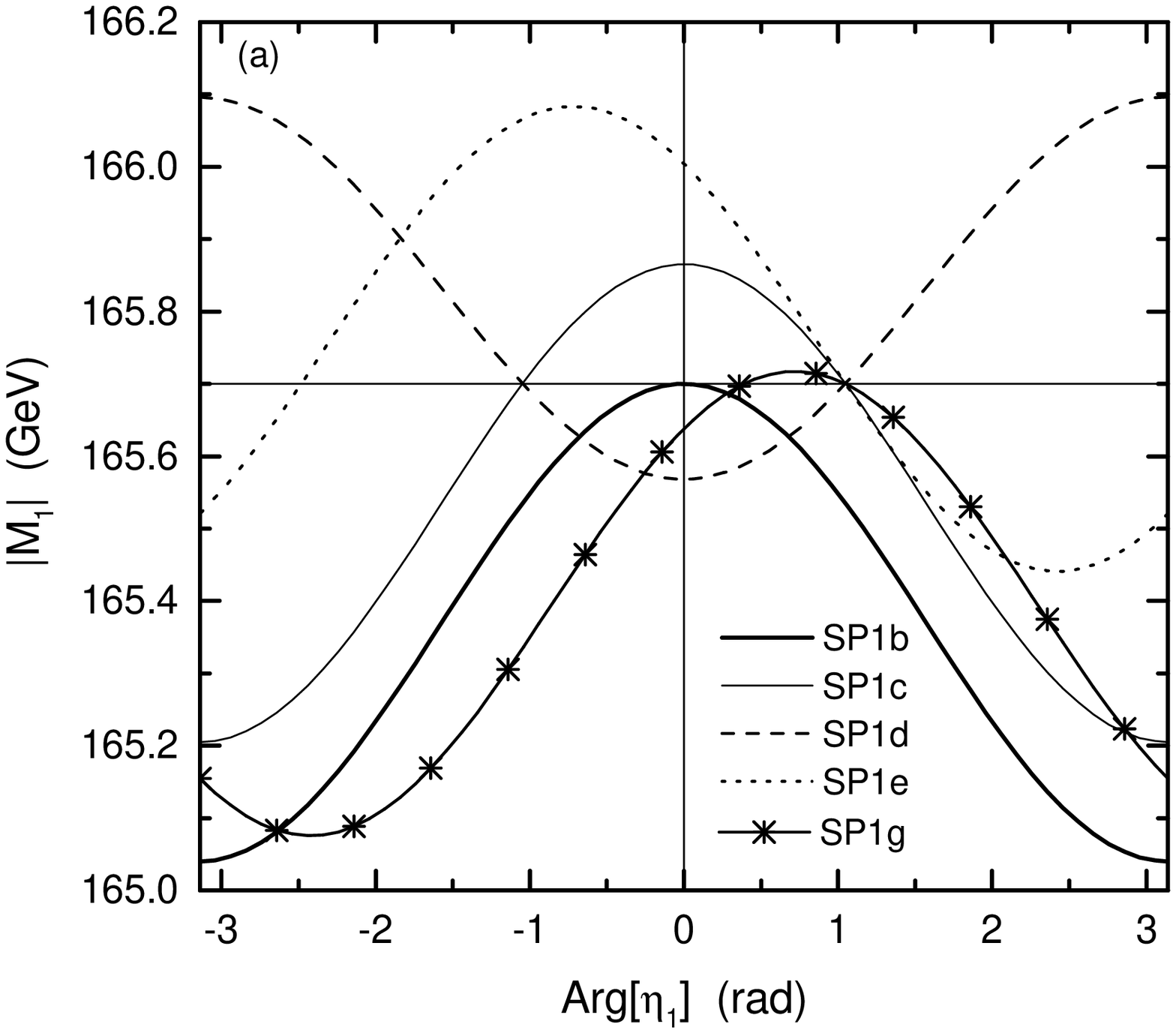,height=7.cm}\hspace{1.0cm}
\epsfig{file=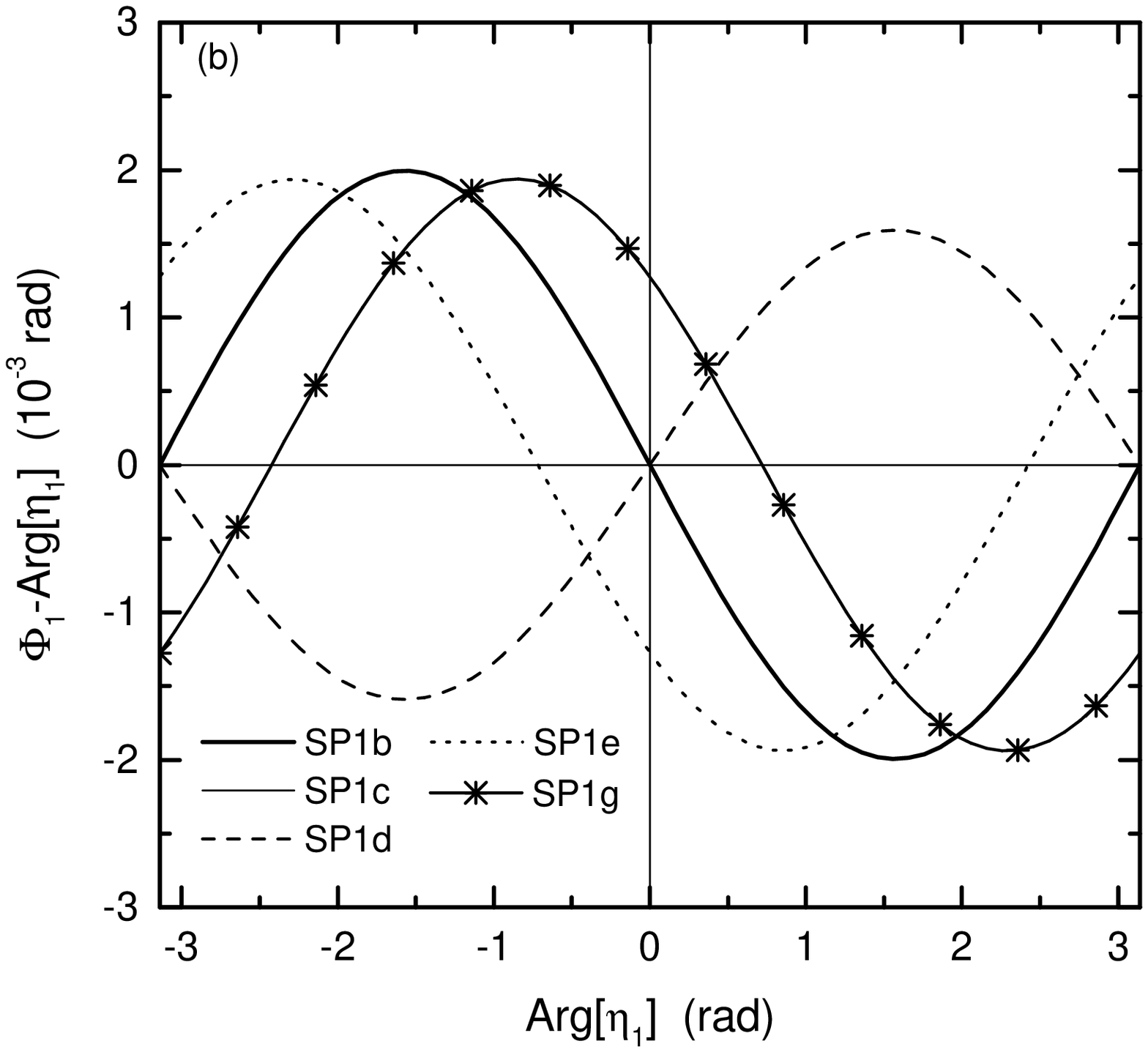,height=7.cm}
\]
\vspace{-0.5cm}
\caption[1]{Reconstruction of $|M_1|$ and $\Phi_1$ in
the CPv1 scenario, in which $M_2$,  $\tan\beta$, $|\mu| $ and
$\Phi_\mu$ and the physical mass of the lightest
neutralino $\tchi_1$
 are assumed as in models $SP1b$, $SP1c$, $SP1d$,
$SP1e$, $SP1g$; see Table 2. The neutralino
CP-eigenphase  $\eta_1$ is allowed to vary. In (b)
the predictions for $SP1b$ and $SP1c$ overlap. }
\label{M1m1-fig}
\end{figure}

If  the  known neutralino
is the lightest one $\tchi_1$,  we present in Fig.\ref{M1m1-fig}a
the reconstructed value of $|M_1|$ calculated from
(\ref{M1comp}),  as a function of the
 phase angle ${\rm Arg } (\eta_1)$. As seen from this
 figure, the reconstructed value of
 $|M_1|$ varies extremely  little around the expected
 value of 165.7 GeV, as  ${\rm Arg } (\eta_1)$
takes  all  possible values.
Under the same conditions, Fig.\ref{M1m1-fig}b indicates that the
reconstructed value of $\Phi_1$ closely follows that of
$\eta_1$.

We conclude therefore that, at least  in these models,
$|M_1|$ can be accurately reconstructed from the physical mass of
the lightest neutralino, even if we know nothing about its
CP-eigenphase $\eta_1$.

\begin{figure}[b]
\vspace*{-1.5cm}
\[
\hspace{-0.5cm}
\epsfig{file=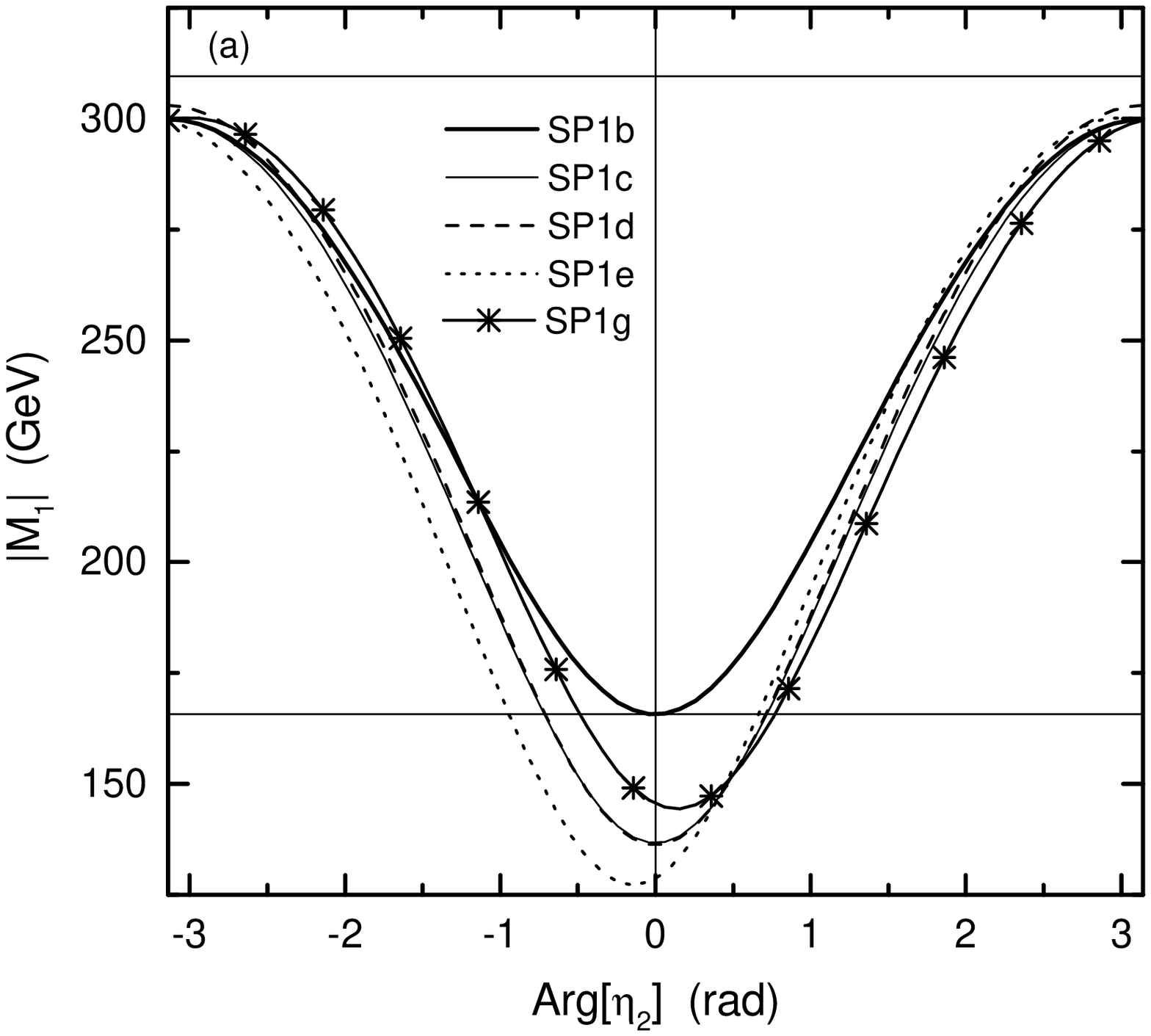,height=7.cm}\hspace{1.0cm}
\epsfig{file=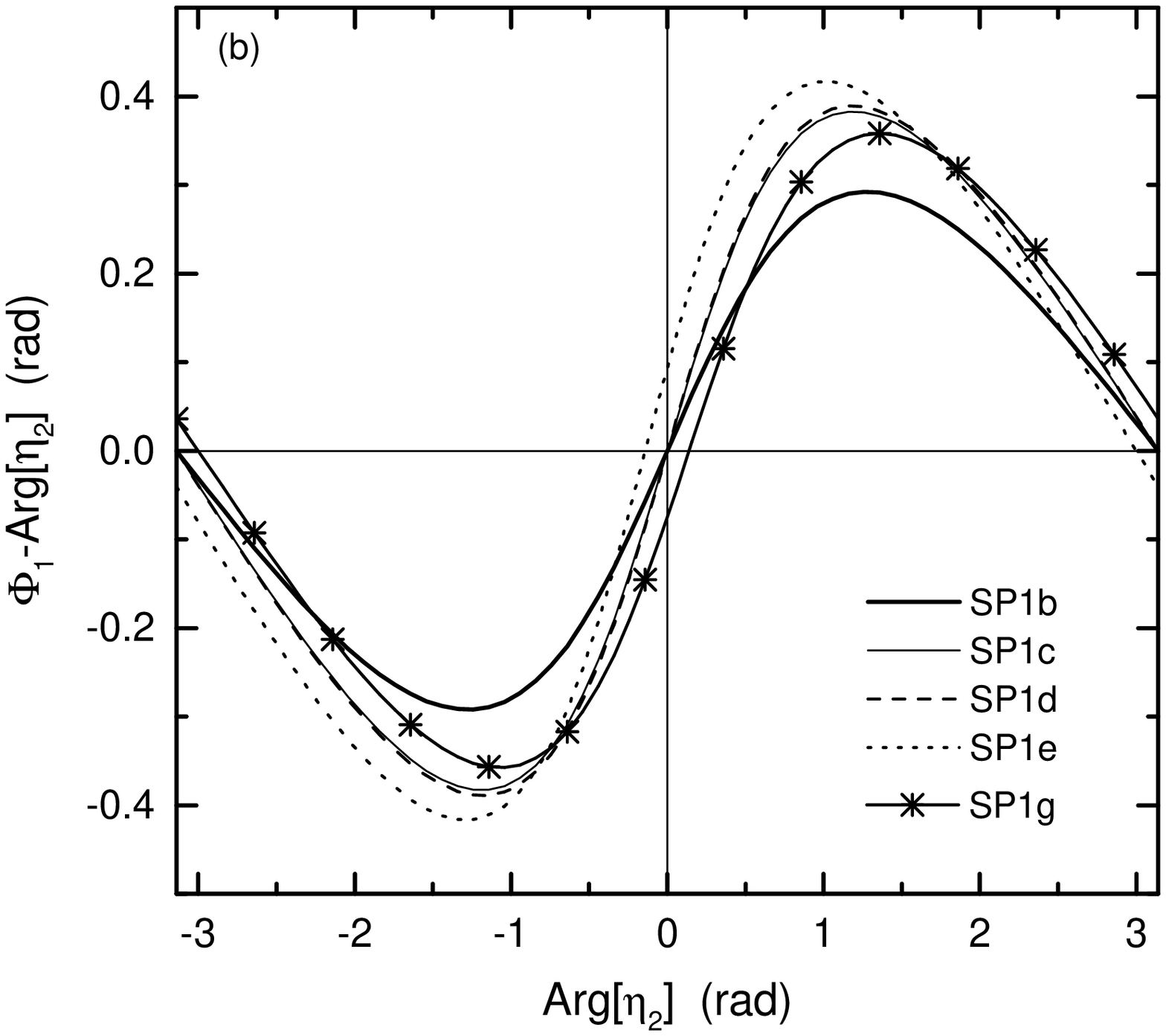,height=7.cm}
\]
\caption[1]{Same as in Fig. 1, but considering instead the
$\tchi_2$ physical mass of the neutralino. The neutralino
CP-eigenphase  $\eta_2$ is allowed to vary. }
\label{M1m2-fig}
\end{figure}

In principle, the same method
could have been also applied in the case
where the known neutralino  mass was $m_{\tchi_2}$.
But, as  it is clearly seen
from Figs.\ref{M1m2-fig}a,b,  the method is now
 not expected to be accurate.

\begin{figure}[p]
\vspace*{-3cm}
\[
\hspace{-0.5cm}
\epsfig{file=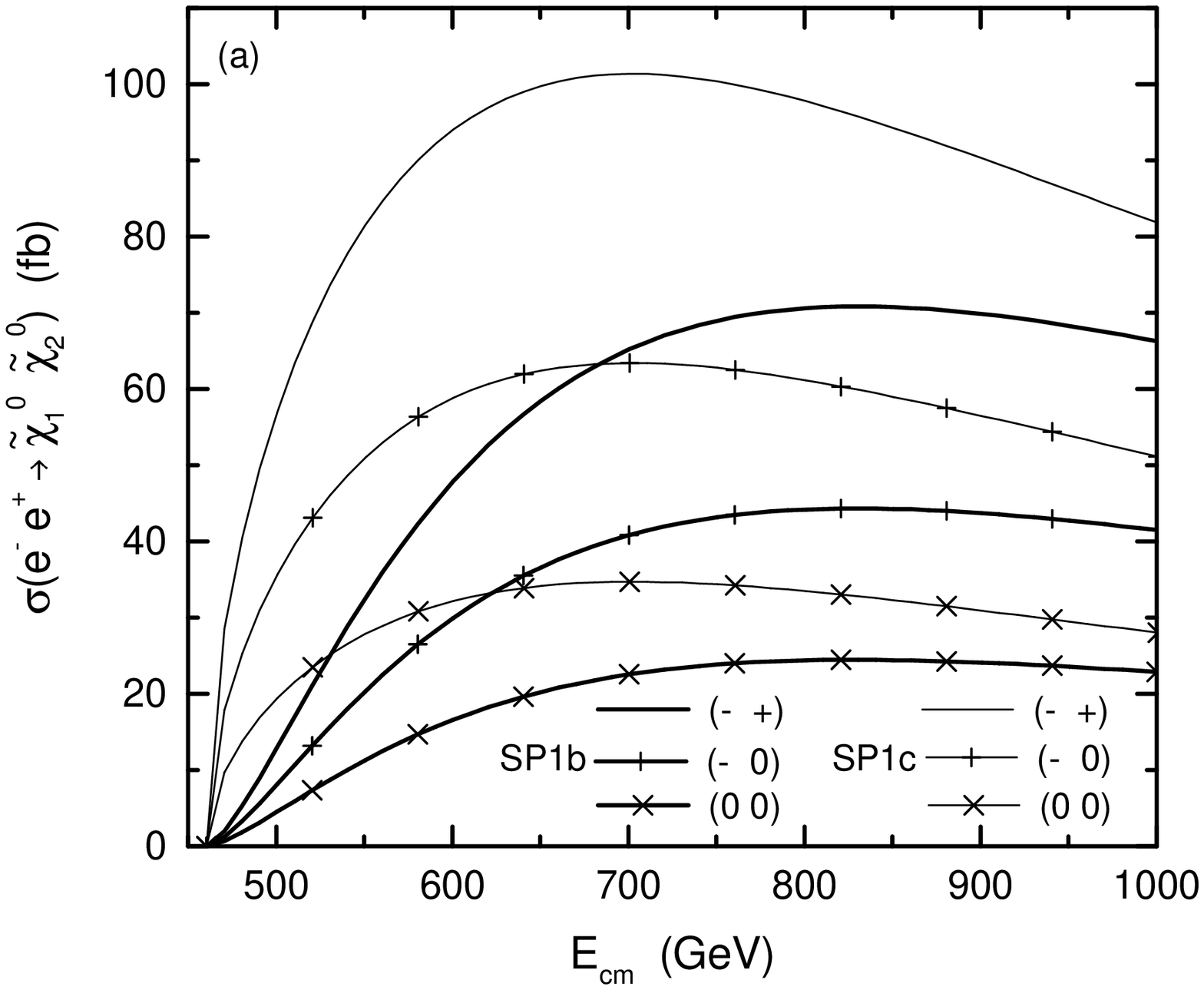,height=8.7cm}\hspace{-1.0cm}
\epsfig{file=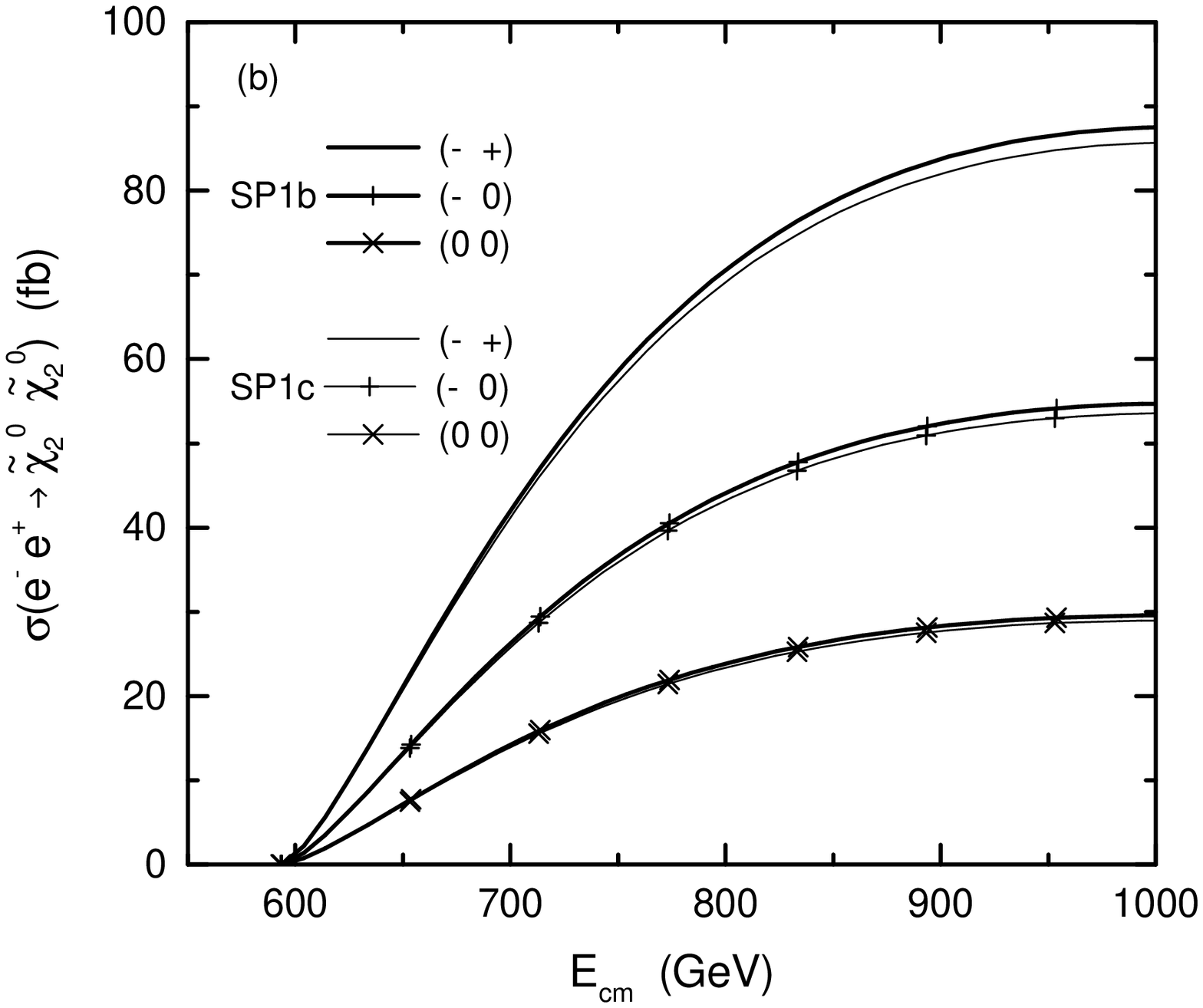,height=8.7cm}
\]
\vspace*{-1.0cm}
\[
\epsfig{file=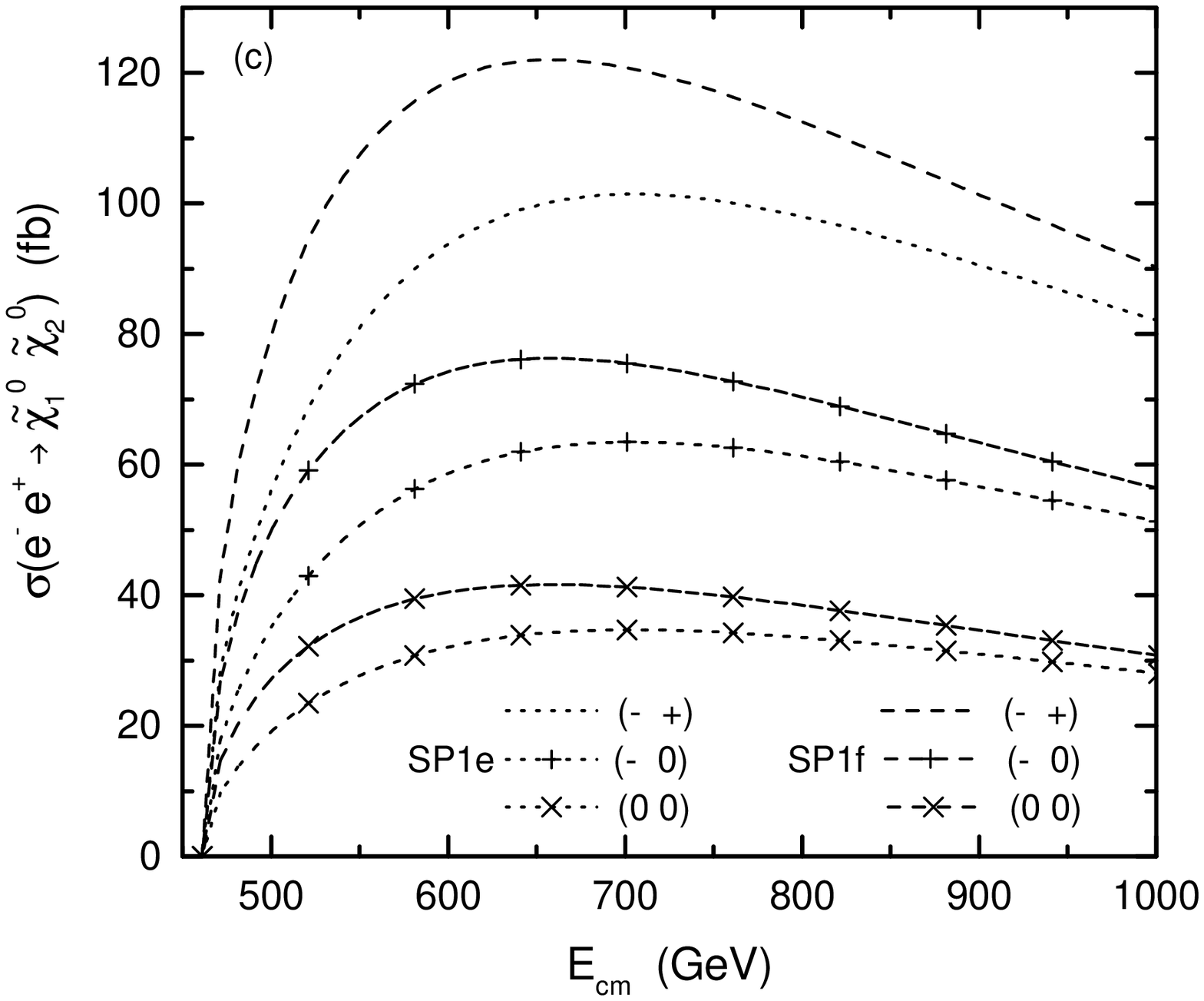,height=9.cm}
\]
\caption[1]{Comparison of the CP-conserving
SP1b ($\eta_1=\eta_2$), and the
CP-violating model SP1c ($\eta_1\neq \eta_2$ and complex) for
$\sigma(e^-e^+\to \tchi_1 \tchi_2)$ (a), and
$\sigma(e^-e^+\to \tchi_2 \tchi_2)$ (b) respectively.
In (c), the CP violating models SP1e and SP1f  are compared for
$\sigma(e^-e^+\to \tchi_1 \tchi_2)$. Polarizations as in
the caption of Fig.\ref{sigma-CPc-fig}a. For clarity, we also
note that the marks on the  (-0) lines,  are actually pluses (+).}
\label{sigma-CPv-fig}
\end{figure}

\vspace{0.2cm}
We next turn to the $\Phi_1$ determination. From
Fig.\ref{M1m1-fig}b, we expect
$\Phi_1 \simeq \eta_1$. For measuring therefore
$\Phi_1$, we need to know the CP eigenphase $\eta_1$ of
the lightest neutralino.
Information on $\eta_1$ could be obtained \eg by looking at
neutralino production in an $e^-e^+$ Linear Collider.
In Figs.\ref{sigma-CPv-fig}(a) and (b) we present the cross
sections for
$\tchi_1 \tchi_2$ and $\tchi_2 \tchi_2$ production respectively.
The results correspond to the models $SP1b$  and $SP1c$,
the parameters of which appear in
Table 2. As expected from (\ref{chiij-thresh},
\ref{chiij-thresh-par}), the dominant $\Delta s^{1/2}$ term in the
threshold region for the $\tchi_1\tchi_2$ production cross section,
should be vanishing in  the CP conserving model $SP1b$
where $\eta_1=\eta_2$, but not in the $SP1c$ case in which
 $\eta_1\neq \eta_2$ and complex.

For $\tchi_2\tchi_2$ production though, the  first non
vanishing threshold term is $\Delta s^{3/2}$.  From
Fig.\ref{sigma-CPv-fig}a,b and an expected luminosity
of $500~fb^{-1}$, we see that the $e^- e^+ \to \tchi_1 \tchi_2$ cross
section measurement can indeed discriminate between the
models  $SP1b$ and\footnote{According to (\ref{z2-CP}-\ref{D-aux}),
 the upper and lower sign  models appearing
 in Table 2 as $SP1c$ (or $SP1d$)
 are characterized by complex conjugate projectors and
 $(\eta_1,~ \eta_2)$ phases. They therefore give identical
 cross sections from (\ref{chiij-dcross}).}
 $SP1c$, which differ essentially only
 for $\Phi_1$ and the neutralino eigenphases.

In the above numerical example we have thus  found
 that if the measured neutralino is
the lightest one; then,
 $|M_1|$ and $\Phi_1$ may be accurately reconstructed.
 In that example, we have also  found that
 $|M_1|\simeq m_{\tchi_1}$ and $\Phi_1\simeq \eta_1$.

\vspace{0.3cm}
\noindent
$\bullet$ The scenario CPv2:\\
In addition to the chargino
parameters $M_2$, $\tan \beta$, ~$\bar \mu$ and $\Phi_\mu$,
we assume here that
we also know the physical masses of  two  neutralinos;
say \eg the masses of the two lightest neutralinos
$m_{\tchi_1}, ~m_{\tchi_2}$.
The CP eigenphases $\eta_1, ~\eta_2$ are taken as
unknown \cite{Kneur-CPv}.

Then  $\bar  M_1\equiv |M_1|, ~ \Phi_1$
and the CP-eigenphases $\eta_1, \eta_2$
can be extracted analytically, up to a two-fold discrete ambiguity .
Indeed applying (\ref{M1comp}) to  the two lightest neutralinos we
obtain
\bq
A_1 \eta_1- A_2\eta_2 = B_2-B_1 ~~, \label{eta1eta2}
\eq
where  $A_1, ~A_2, ~B_1, ~B_2$ are defined in (\ref{AjBj-CP}).
Imposing then the condition $|\eta_{1,2}|=1$ in (\ref{eta1eta2})
and defining $\Delta B \equiv B_2 - B_1$,  we obtain the
constraint\footnote{It is equivalent to Eq.(3.15) of
 \cite{Kneur-CPv}. We thank Gilbert Moultaka for
 pointing this out to us.}
\bq
\label{deltap}
\tilde \Delta = 4 A_1^2 A_2^2-(A_1^2+ A_2^2-|\Delta B|^2)^2 \ge 0
~~ ,
\eq
together with
\bq
\label{eta12}
\eta_{1,2}=\frac{A_1^2-A_2^2 \pm |\Delta B|^2+i \epsilon
  \sqrt{\tilde \Delta}}{2 ~ \Delta B^* ~
A_{1,2} }
\eq
and
\bq
\bar  M_1 \exa =\frac{A_1^2-A_2^2+(B_1+B_2)
\Delta B^*+i \epsilon  \sqrt{\tilde \Delta}}{2 ~ \Delta B^*} ~~,
\label{M1comp-CPv2}
\eq
where $\epsilon= \pm 1$
expresses the aforementioned  two-fold ambiguity of the solution.
In the context  of this  scenario,
(\ref{eta12}) fixes the CP eigenphases of the two lightest
neutralinos, and  (\ref{M1comp-CPv2})
determines $(\bar M_1, ~\Phi_1)$, up to a two-fold ambiguity.

As an example we consider the models $SP1e$ and $SP1g$ of Table 2,
considering
as input ($M_2$, $\tan \beta$, ~$\bar \mu$, $\Phi_\mu$)
together with the
physical masses of the two lightest neutralinos. As expected,
(\ref{M1comp-CPv2}, \ref{eta12}) reproduce
the $\bar M_1$, $\Phi_1$,  $\eta_1$ and $\eta_2$
results quoted in Table 2
for $SP1e$ and $SP1g$ respectively; and in addition,
due to the above ambiguity,  they also respectively allow the
models $SP1f$ and $SP1h$ shown in the same Table.

Exactly the same statements  could be made  when  applying the above
procedure to the  upper signs  version of the model
$SP1c$ (or $SP1d$) in Table 2; and obtaining,
due to the aforementioned ambiguity, also
 the lower signs version  of the same model.

As seen from  Table 2,  the  models in each of the pairs
 ($SP1e$, $SP1f$)
 and  ($SP1g$, $SP1h$)  have the same values for
 $M_2$, $\tan\beta$, $|\mu|$, $\Phi_\mu$
 and the physical masses of both  charginos and the
  two lightest neutralinos. They have also  almost the same
  values for  $|M_1|$. Their main difference is on the phase angles
   $\Phi_1, ~{\rm Arg}(\eta_1), ~{\rm Arg}(\eta_2)$,
   which are found to be of opposite sign, in
  each of the above pairs. Apart from this, their only other
  difference is on  the  higher neutralino masses
  and CP   eigenphases.

  An efficient   way to discriminate  between  models like \eg
   $SP1e$ and $SP1f$,  is by looking at
  $\sigma (e^- e^+\to \tchi_1 \tchi_2)$. These cross sections
   are presented in Fig.\ref{sigma-CPv-fig}c for polarized beams.
  According to (\ref{chiij-dcross}-\ref{ZseR-int})
  and the reduced projectors of
  (\ref{z2-CP}-\ref{z4-CP}), the difference between  the cross sections
  for these   models   solely comes from the
  differences between     $\eta_1$ and $\eta_2$ given in Table 2.
  Thus, for polarized beams with longitudinal polarizations
  $\lambda_1=-0.85$,   $\lambda_2=0.6$, the cross
   section difference between the
   two models may reach the level of 20  fb;
   which means a difference  of about ten thousands of events.
The discrimination between such models should  be amply
possible at a Linear Collider.

\vspace{0.3cm}
\noindent
$\bullet$ The scenario CPv3:\\
This scenario is a weak version of CPv2. We assume in it,  that
the first quantities to be measured will be
the physical masses of the two lightest
neutralinos,  the mass of the lightest chargino $\tchic_1^+$,
and in addition to it, the chargino
mixing angle $\phi_L$ and  $\tan \beta$
   \cite{Zerwas1, Kneur-CPv}.

The motivation for this scenario is based on the observation
that a  study of the
$e^+ e^- \rightarrow \tchic_1^+ \tchic_1^-$
cross section for polarized beams,
should be sufficient to extract  $\tan \beta$ and the
mixing angle $\phi_L$, even if it would be impossible
to experimentally determine the mass of the heavier chargino
\cite{Zerwas1, Kneur-CPv}. Thus, $m_{\tchic^\pm_2}$
 is not assumed known.

The relevant formulae in this scenario are
still (\ref{deltap}, \ref{eta12}, \ref{M1comp-CPv2})
reconstructing the neutralino
eigenphases and complex $M_1$, in terms of $\tan\beta$ and
 $(M_2, ~\mu)$. But now $(M_2, ~\mu)$ should also be
 constructed from  the lightest  chargino
 mass and $\phi_L$. To this end we note that the chargino mass matrix
 gives \cite{Abdel-CPv1}
\bqa
|\mu|&=&\sqrt{ \frac{m_{\tchic_2^\pm}^2
[1+\cos (2\phi_L)]+m_{\tchic_1^\pm}^2[1-\cos (2\phi_L)]-4 \mwd
\cbeta^2}{2}} ~~ ,\nonumber \\
M_2&=&\sqrt{ \frac{m_{\tchic_2^\pm}^2
[1-\cos (2\phi_L)]+m_{\tchic_1^\pm}^2 [1+\cos (2\phi_L)]
-4 \mwd \sbeta^2}{2}} ~~, \label{M2mugmu-CP}
\eqa
and
\bqa
\label{phimu-rec}
\cos\Phi_\mu  &=& 1-\frac{m_{\tchic_1^\pm}^2 m_{\tchic_2^\pm}^2
-(M_2 |\mu|-\mwd
\s2beta)^2}{2 \mwd M_2 |\mu|
\s2beta} ~~,  \nonumber \\
\sin \Phi_\mu &=& \epsilon^{'} \sqrt{1-\cos^2 \Phi_\mu } ~~,
\eqa
involving   an additional two-fold ambiguity on
the sign of $\Phi_\mu$, which is  described by
$\epsilon'=\pm 1$,  \cite{Kneur-CPv}.

To  reconstruct   $(|\mu|, ~M_2, ~\Phi_{\mu})$
from (\ref{M2mugmu-CP}-\ref{phimu-rec}), we
first need to constrain or  determine  $m_{\tchic_2^\pm}$.
A  very strong such constraint  may be
obtained by  imposing on (\ref{phimu-rec}) the requirements
$|\cos\Phi_\mu|\leq 1$, $|\sin\Phi_\mu| \leq  1$,   \cite{Kneur-CPv}.
To illustrate this  we consider again the model $SP1e$
of Table 2, for which  the  chargino masses are
$m_{\tchic_1^\pm}=296.9 ~  GeV$,
$m_{\tchic_2^\pm}= 516.17~ GeV $ and
$\cos (2\phi_L)=-0.77 $.

If now, within the philosophy of the CPv3
 scenario, we invert the logic and
 start from the   $SP1e$ values for
$m_{\tchic_1^\pm}$, $\tan \beta$,  $\cos (2\phi_L)$ and
 $m_{\tchi_1},~ m_{\tchi_2}$; while
 treating   $m_{\tchic_2^\pm}$, as well as all other $SP1e$
 parameters,  as unknown;  then consistency of (\ref{phimu-rec})
 implies the rather strong constraint
\[
507.2 ~{\rm GeV} ~ \lsim ~ m_{\tchic_2^\pm}~ \lsim  ~517.7 ~
{\rm GeV } ~~.
\]
This  constraint  is generated from the chargino sector alone,
and it is  independent of the  $m_{\tchi_1}$ and
 $m_{\tchi_2}$ values. It is further strengthened if
we also  demand
that $m_{\tchic_2^\pm}$ should be such, that the
induced values for  $|\mu|$ and $M_2$ from (\ref{M2mugmu-CP}),
satisfy the constraint (\ref{deltap}). Indeed, for the
 same input parameters as above, and taking also  into
  account the information on the neutralino masses, we then  find
 the very tiny interval
\bq
515.8 ~{\rm GeV} ~ \lsim ~ m_{\tchic_2^\pm}~ \lsim  ~516.9 ~
{\rm GeV } ~~, \label{SP1emchic2}
\eq
independently of the $\epsilon$, $\epsilon '$ ambiguities.
The   corresponding
intervals from (\ref{M2mugmu-CP}) are
\[
495.6 ~{\rm  GeV}  \lsim ~  |\mu| ~ \lsim  496.5 ~{\rm GeV}
~~~~ , ~~~~~
309.5 ~{\rm GeV} \lsim ~  M_2 ~ \lsim 309.7  {\rm GeV }
~~.
\]
$\Phi_\mu$ is then restricted by
(\ref{phimu-rec}) to the $\epsilon'~$-dependent interval
\[
0.63 ~{\rm  rad}  \lsim  \epsilon ' \Phi_\mu \lsim 0.87 ~{\rm rad}
~~,
\]
while  (\ref{M1comp-CPv2}) gives
\bq
165.69 ~{\rm  GeV}  \lsim   \bar  M_1  \lsim  165.72 ~{\rm GeV}
~~~~ , ~~~~~
1 ~{\rm  rad}  \lsim  \epsilon '~ \Phi_1 \lsim 1.1 ~{\rm rad}
~~, \label{SP1eM1}
\eq
for $\epsilon \epsilon '=-1$, and
\bq
165.99 ~{\rm  GeV}  \lsim   \bar  M_1  \lsim  166.72 ~{\rm GeV}
~~~~ , ~~~~~
-1.34 ~{\rm  rad}  \lsim  \epsilon' ~\Phi_1 \lsim -1.31 ~{\rm rad}
~~, \label{SP1fM1}
\eq
for  $\epsilon \epsilon '=1$.

We note that the appearances of $\epsilon$ and $\epsilon'$ above
and in (\ref{eta12},  \ref{M1comp-CPv2}),
indicate a four fold ambiguity. If $\epsilon'=1$ is selected, then
(\ref{SP1eM1}) and (\ref{SP1fM1}) correspond respectively to the
 models $SP1e$ and $SP1f$, in Table 2.
 On the other hand, the solution  $\epsilon'=-1$
 combined with  $\epsilon= 1$ or $\epsilon= -1$,
 allows two additional model possibilities,
 in which the reduced projector elements and the CP
 eigenphases for the two lightest neutralinos will be
 complex conjugate to those of the models $SP1e$ and $SP1f$
 respectively; compare (\ref{eta12},
 \ref{M1comp-CPv2}). For clarity, we  call these later models
 $SP1e'$ and $SP1f'$, respectively\footnote{They should not
  be confused with  $SP1g$ and $SP1h$ of Table 2,
  in which the masses of the  two lightest neutralino are
   different.}. \\

The above constraints are derived assuming that
we only know the physical masses of the two lightest
neutralinos,  the mass of the lightest chargino $\tchic_1^+$,
$\tan \beta$ and the mixing angle $\phi_L$.
It turns out that we can go beyond  this, and fully determine
the mass of $\tchic_2^\pm$, even if
(as in   $SP1e$), it is too heavy to be produced
on-shell in a 500 GeV LC.
Such a determination  may \eg be done
if  we also measure
 $\sigma (e^- e^+ \to \tchi_1 \tchi_2)$,
 preferably for polarized beams.

Because of the aforementioned properties of the RPE
and $\eta_{1,2}$,  the models in each of the
pairs ($SP1e$, $SP1e'$) or ($SP1f$, $SP1f'$)
give identical results for
$\sigma (e^- e^+ \to \tchi_1 \tchi_2)$.
Thus, measurement of this cross section may only be used to
 discriminate between the   pairs ($SP1e$, $SP1e'$) and
  ($SP1f$, $SP1f'$), corresponding to
   $\epsilon \epsilon'= -1$
and  $\epsilon \epsilon'= 1$, respectively.

Thus, using (\ref{M2mugmu-CP}, \ref{phimu-rec}, \ref{eta12}),
 and the reduced projector
 elements $p_{1\alpha}$, $p_{2\beta}$  evaluated
from (\ref{z2-CP}-\ref{z4-CP}),  the cross
section $\sigma (e^- e^+ \to \tchi_1 \tchi_2)$ is calculated from
(\ref{chiij-dcross}).
For a given energy $\sqrt{s}$ and
polarizations $(\lambda_1, \lambda_2)$, this cross section only
depends on the chargino mass $m_{\tchic_2^\pm}$, allowed to
vary in the interval specified by  $|c_{\mu}|, |s_{\mu}| \le 1$ and
$\tilde \Delta \ge 0$. The mass $m_{\tchic_2^\pm}$ is
obtained by comparison with the experimental cross
section.

As an illustration of how well we can reproduce the specific
"experimental" value of $m_{\tchic_2^\pm}=516.17~GeV$
for  the  above  $SP1e$-generated   parameters,
 we call  "experimental", the  cross
 section\footnote{As already stated, this cross section
 is identical for both $SP1e$ and $SP1e'$ models.}
$\sigma (e^- e^+ \to \tchi_1 \tchi_2)_{\rm exp} \simeq 54.5 ~{\rm fb}$
  for $m_{\tchic_2^\pm}=516.17~GeV$, $\epsilon \epsilon'=-1$,
  $\sqrt{s}=500 ~ GeV$
  and  $e^\mp$-polarizations $(\lambda_1, \lambda_2)=(-0.8,0.6)$.
   The "theoretical" cross
section is defined  for varying $m_{\tchic_2^\pm}$
in the range of (\ref{SP1emchic2}), considering the two
possibilities  $\epsilon \epsilon'= -1$ and $\epsilon \epsilon'= 1$
corresponding to ($SP1e,~ SP1e'$) and ($SP1f,~ SP1f'$)
respectively.

\begin{figure}[t]
\vspace*{-3cm}
\[
\epsfig{file=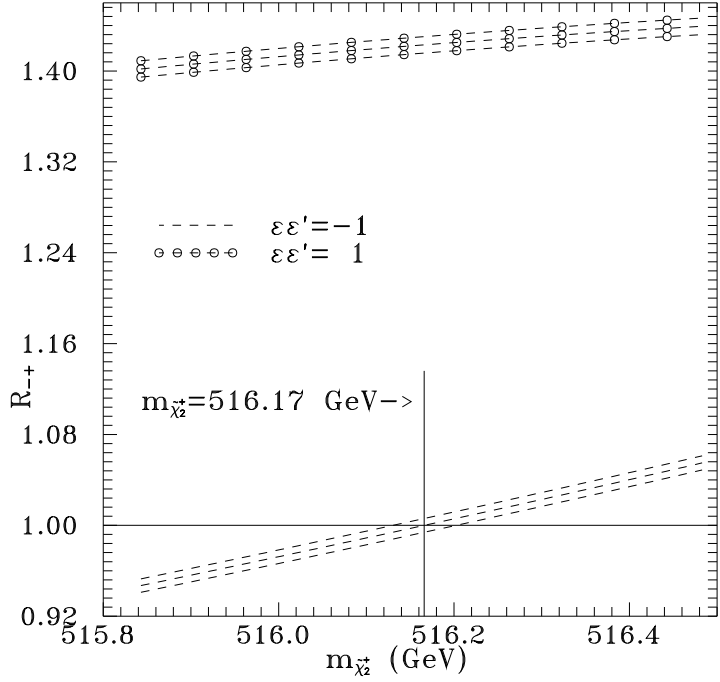,height=9.0cm}
\]
\vspace*{-0.5cm}
\caption[1]{Reconstruction of $m_{\tchic_2^\pm}$ in
the  CPv3 scenario, for models ($SP1e, ~SP1e'$).
Only  the ambiguity between  $\epsilon \epsilon'=1$,
($SP1f, ~SP1f'$ ) and $\epsilon \epsilon'=- 1$, ($SP1e, ~SP1e'$)
 may be lifted. As input we use
  $m_{\tchic_1^\pm}=296.9~GeV$, $\cos(2\phi_L)=-0.77$, and the
  two lightest neutralino masses and $\tan\beta$   given
   in Table 2. The $\tchic_2^\pm$ "experimental"
     value to be reproduced is  $m_{\tchic_2^\pm}= 516.17 ~{\rm GeV}$.
     The LC energy is 500 GeV and $\L =500 ~fb^{-1}$.
     The middle dash (dash-circles) lines
     indicate the central ${\cal{R}}_{-+}$-value, while the upper and
     lower lines describe  the 1 SD changes.}
\label{Ratio-fig}
\end{figure}

In Fig.\ref{Ratio-fig}, we  display the ratio of these cross
sections
\bq
{\cal R_{-+}}= \frac{\sigma (e^-e^+\to \tchi_1\tchi_2)_{\rm th}}
{\sigma (e^-e^+\to \tchi_1\tchi_2)_{\rm exp}} ~~ ,
\label{Rmp}
\eq
as a function of  $m_{\tchic_2^\pm}$.

For   a Linear Collider with an integrated
luminosity of about $\L= 500 ~ fb^{-1}$; the 1 SD error to
${\cal R}_{-+}$ is
\bq
\label{delta-Rmp}
\delta {\cal R}_{-+} \simeq \sqrt{\frac{ {\cal R}_{-+}}
 {\sigma(e^- e^+ \to \tchi_1 \tchi_2)_{\rm exp} ~\L }} ~~.
\eq
Using this, we also present in Fig.\ref{Ratio-fig}
the $\pm 1 ~{\rm SD}$ results
 ${\cal R}_{-+}~ \pm~ \delta {\cal R}_{-+}$, for
 the two distinct cases $\epsilon \epsilon' =-1 $ and
 $\epsilon \epsilon' =+1 $. As seen in  Fig.\ref{Ratio-fig},
 the possibility $\epsilon \epsilon' =+1 $
 (corresponding to the pair  $SP1f,~ SP1f'$)
   is   ruled out by many standard deviations;
   while for  $\epsilon \epsilon' =-1 $ we find
$m_{\tchic_2^\pm}= 516.20 \pm 0.04 $, in consistency with
models $SP1e$ and  $SP1e'$.

After $m_{\tchic_2^\pm}$ is thus fixed;  then the  choice
$\epsilon'=1$, together with
(\ref{M2mugmu-CP}-\ref{phimu-rec},
\ref{deltap}, \ref{eta12}, \ref{M1comp-CPv2})
determines $|\mu|$, $\Phi_\mu$, $M_2$, $|M_1|$, $\Phi_1$,
and the CP eigenphases of the two neutralinos we have started with
$\eta_1, ~\eta_2$, to the values expected for the $SP1e$ model.
Of course the $\epsilon'$ ambiguity is not lifted, and an
alternative set of results is allowed corresponding to $SP1e'$.

The same procedure could be applied also for a c.m. energy of
$\sqrt{s}=800 ~ {\rm GeV}$. For the same polarizations
$(\lambda_1, \lambda_2)=(-0.8,0.6)$, we again find that
$\epsilon \epsilon' =+1$ is strongly ruled out;
while the reconstruction of the chargino mass
$m_{\tchic_2^\pm}$ for $\epsilon \epsilon' =-1 $ gives
$m_{\tchic_2^\pm}= 516.2 \pm 0.1 $. It is interesting to note that
the reconstruction error   is larger in this case.

A similar procedure could also be applied to the pair of models
labeled  $SP1c$ (or $SP1d$). In this particular  case, where the
chargino sector is CP conserving, we should only have the
$\epsilon$ ambiguity; not the $\epsilon'$ one, \cite{Zerwas1}.
But the measurement of
$\sigma (e^- e^+ \to \tchi_1 \tchi_2)$ will not be able to lift
it. So finally again one ambiguity remains which is expressed
by the upper and lower signs in each of these pairs of models.

\section{Conclusions}

In this paper we have first addressed  the problem of constructing
analytical expressions for the diagonalization of the most general
 complex symmetric neutralino mass matrix. The motivating idea was
 to extend to complex SUSY parameters, the approach of \cite{GLMP},
 which was hitherto applied only to the real SUSY parameter case.
  A very nice feature of this approach is  that it
can be straightforwardly   extended  to theories
containing an arbitrary number of neutralinos.

 The diagonalization   starts from the observation that
 neutralino physical observables can only depend on their
 projectors (or density matrices);
  and what we have here defined as the neutralino
  pseudo-projectors,   enforced by their Majorana nature.

  In the next step, we introduce for each physical neutralino
  $\tchi_j$,  its CP-eigenphase $\eta_j$,
   and its "reduced projector elements"
   $p_{j\alpha}$,  which are complex numbers.
 The projectors and pseudo-projectors, as well as the
 diagonalization matrix, can be expressed in terms of them; and
  all may subsequently be calculated using   Jarlskog's formula
  (\ref{Pj-Jarlskog})   together with
  (\ref{P-pj}-\ref{Pj11-pj}, \ref{eqeta0}).  Analytic
expressions for calculating the physical masses are given in
Appendix A.

 The end result  is that all physical observables
related to any specific neutralino, can be fully described in
terms of its CP eigenphase and its reduced projector elements.
The ensuing formulae
  constitute one of the basic contents  of this paper.
  Their   importance is further emphasized by noting
  that they can be directly generalized to
  any non-minimal SUSY theory, involving any number of
  neutralinos with real or complex couplings.

As an example, we have presented the tree-level expression for
 ${\rm d}\sigma(e^-e^+\to \tchi_i \tchi_j)/{\rm d}t $
 in the  MSSM case; in which there are three  reduced
 projector elements,
and of course one CP eigenphase, for each neutralino.
Arbitrary beam polarizations are used.

 The above formulae are quite simple and easy to insert in a
 numerical code. Thus, they  should be useful in any
 calculation involving either on-shell
 or virtual neutralinos. Since all  masses and mixings
 are determined  together  at the same  scale and
 level of approximation; it is  guaranteed that
 no   spurious enhancements  or
 lack of divergence cancellations in  neutralino  loops,
 may creep in.\\

 The next very  important set of formulae  of the present paper
 appears in (\ref{M1c}-\ref{mchic}) and
 expresses the various MSSM  parameters in terms  of
 the three reduced projector  elements
  $p_{j \alpha}$ ($\alpha=2-4$) for
  $\tchi_j$,  its  mass $m_{\tchi_j}$,
 and its CP eigenphase $\eta_j$.
 These relations are subsequently used in various applications.

 The first  concerns   the CP conservation in
 the chargino-neutralino sector and may be stated  as: ~
  CP conservation is equivalent to the reality of
 the  reduced projector  elements referring to  any
 given neutralino state. But if this happens, then
 it is  immediately concluded
 that all reduced projector elements for all neutralinos, and all
 CP eigenphases, are in fact  real.

   As a second set of applications of (\ref{M1c}-\ref{mchic}) and
   their inverse (\ref{eqeta0}-\ref{eqeta4}), we have studied
     various scenarios
   of reconstructing the SUSY parameters; under various conditions
   concerning the acquirement of experimental knowledge
   in the chargino and
   neutralino sectors.  To study the efficiency of our formalism,
   we have in fact considered  the same scenarios
   as in  \cite{Kneur-CPc, Kneur-CPv, Zerwas1}.
   In all    cases,  analytical expressions are  given which
   disentangle  the unknown SUSY parameters. There is never
    need to do this numerically \cite{Zerwas1, Kneur-CPv}.
      Ways to handle the  various construction ambiguities are also
   discussed.

    More explicitly, in the real SUSY parameter CP conserving
    case,      we considered the scenario {\cal S}1,
   in which  $M_2,~ \mu$, $\tan\beta$, the lightest
   physical neutralino mass  and (partly at least) its CP eigensign
   are known; and {\cal S}2, in which
   measurements of only
   the masses of the lightest chargino and the two lightest
   neutralinos, are assumed.

   For the complex parameters case, we have also considered
   scenarios in which  chargino measurements are assumed to have
   already  determined $M_2, ~|\mu|, ~\Phi_\mu$ and $\tan\beta$;
   while from the neutralino sector only
    the physical mass and CP-eigenphase of
   the lightest neutralino (CPv1), or  just  the physical masses of the
   two lightest neutralinos (CPv2), are assumed known.

   As already said,  analytical formulae have in all
   cases been derived  which
   disentangle  the unknown SUSY parameters.

In addition to them,  we have also considered
the  complex parameter CPv3 scenario,
in which only $\tan\beta$, the lightest chargino mass,
  its mixing angle $\phi_L$,  and the
  physical masses of the two lightest neutralinos,
  are assumed known \cite{Kneur-CPv, Zerwas1}. Then, explicit
  expressions are constructed  which may be used to determine
  $m_{\tchic^\pm_2}$ from $\sigma(e^-e^+\to \tchi_1 \tchi_2)$
  measurements; even if the energy is not sufficient to produce
  $\tchic_2^\pm$. Once, $m_{\tchic^\pm_2}$ is determined;
  expressions for finding $|\mu|$, $\Phi_\mu$, $M_2$, $|M_1|$,
  $\Phi_1$, as well as the CP eigenphases of the two
  lightest neutralinos, are also given.

\vspace{0.2cm}
In all previous scenarios, some knowledge of the chargino sector
was assumed, and only the total neutralino production cross section
at an LC was used. Since in many models
$m_{\tchi_2}\sim m_{\tchic^+_1}$, it may
turn out that the charginos will be too heavy to be produced,
and that  the
neutralino sector will be the first to be studied.
In such a case,  the differential cross
section $ {\rm d} \sigma(e^- e^+ \to \tchi_i \tchi_j)/{\rm d} t $,
together with the physical masses of the two lightest neutralinos,
will  be the only means we might have in order to reconstruct
the neutralino parameters.
We plan  to explore such scenarios in the near future.

\vspace{0.5cm}
\noindent
{\bf \Large Acknowledgements.}\\
We are grateful to Abdelhak  Djouadi, Jean-Loic Kneur and Gilbert
Moultaka for many helpful communications.

\clearpage
\newpage

\renewcommand{\theequation}{A.\arabic{equation}}
\renewcommand{\thesection}{A.\arabic{section}}
\setcounter{equation}{0}
\setcounter{section}{0}

{\large \bf Appendix A:
The neutralino masses and projectors in the presence of
the CP violating phases:}\\

Here we present the formulae for the analytic determination of the
neutralino masses. Analogous results have already appeared in
\cite{Zerwas1}. Nevertheless we give them here also, following the
same formalism as in the real SUSY parameter
case treated in  \cite{GLMP}.

Using (\ref{Y-matrix}) we find
\bq
Y^\dagger Y=  \left ( \matrix{ \bar  M_1^2+\mzd \swd  & -\mzd \sw
\cw  & y_{13} & y_{14} \cr - \mzd \sw \cw & M_2^2+ \mzd \cwd &
y_{23}& y_{24} \cr y_{31}& y_{32}&\bar \mu^2+\mzd \cbeta^2 & -\mzd
\sbeta \cbeta \cr y_{41} & y_{42} & - \mzd \sbeta \cbeta  &
\bar \mu^2+\mzd \sbeta^2  \cr}\right ) ~~ ,
 \label{YdagY-matrix}
 \eq
where
\bqa
y_{13}&=&y_{31}^*=-\mz \sw (\bar  M_1 \cbeta \exma+\bar \mu \sbeta
\exb) ~~,  \\
y_{14}&=& y_{41}^*=\mz \sw (\bar  M_1 \sbeta \exma+\bar \mu \cbeta
\exb) ~~, \\
y_{23}&=& y_{32}^*=\mz \cw (M_2 \cbeta+\bar \mu \sbeta \exb)
~~ , \\
y_{24}&=& y_{42}^*=-\mz \cw (M_2 \sbeta+\bar \mu \cbeta \exb)
~~ .
\eqa
The characteristic equation associated to
$Y^\dagger Y$ determine the physical neutralino masses
$m_{\tchi_j}$:
\bq
 m_{\tchi_j}^8- A  m_{\tchi_j}^6 +B  m_{\tchi_j}^4-C  m_{\tchi_j}^2+D=0 ~~,
 \label{characteristic}
\eq with \cite{Abdel-char}
\bqa
A &=& \bar  M_1^2+M_2^2+2 (\bar \mu^2+\mzd) ~~ ,\nonumber \\
B &=& -2 \bar \mu \mzd \s2beta (\bar  M_1 \swd \cab+M_2 \cwd \cb)+(\bar  M_1 M_2)^2
\nonumber \\ && +2 \bar \mu^2 (\bar  M_1^2+M_2^2)+2 \mzd (\bar  M_1^2 \cwd+M_2^2
\swd)+ (\bar \mu^2+\mzd)^2   ~~ , \nonumber \\
C & =&\bar \mu^2 \mz^4 \s2beta^2-2 \bar  M_1 \bar \mu \mzd (M_2^2+\bar \mu^2) \swd \s2beta \cab- 2 M_2
\bar \mu (\bar  M_1^2+\bar \mu^2) \mzd \cwd \s2beta \cb \nonumber \\
&&+\bar  M_1^2 (2 M_2^2 \bar \mu^2+(\bar \mu^2+\mzd \cwd)^2)+M_2^2 (\bar \mu^2+\mzd \swd)^2+ 2 \bar  M_1
M_2 \mz^4 \swd \cwd \ca  ~~,\nonumber \\
D &=& (\bar  M_1 M_2\bar \mu^2)^2+
\mz^4\bar \mu^2
\s2beta^2 (\bar  M_1^2 \cw^4+M_2^2 \sw^4+2
\bar  M_1 M_2 \swd \cwd \ca)
\nonumber \\
 && -2 (\bar  M_1 M_2 \bar \mu^3) \mzd \s2beta (\bar  M_1 \cwd \cb+M_2
\swd \cab)
 ~~ , \label{ABCD-par}
\eqa
where $\s2beta=\sin(2\beta)$, $\ca=\cos \Phi_1$,
 $\cb=\cos \Phi_\mu$ and
$\cab=\cos(\Phi_1+ \Phi_\mu)$.

The  four real and positive solutions of
(\ref{characteristic}) determining $m^2_{\tchi_j}$ are  \cite{GLMP}.
\bqa
 m^2_{\tchi_j}&= & \frac{1}{2}
\Big \{ \frac{A}{2}-2 E \pm \sqrt{\Big (\frac{A}{2}-2
E\Big )^2- 4 \Big (\frac{B}{6}+2 \theta_N +F\Big )} \Big \} ~~ ,
\nonumber \\
&=& \frac{1}{2} \Big \{ \frac{A}{2} +2
E  \pm \sqrt{\Big (\frac{A}{2}+2 E\Big )^2- 4 \Big
(\frac{B}{6}+2 \theta_N - F\Big )} \Big \} ~~ ,
\label{physical-masses}
\eqa
where
\bq
E=\frac{1}{4} \left (A^2-\frac{8B}{3}+16\theta_N \right)^{1/2}
~~~,~~~ F=\frac{1}{4E}\left (C-\frac{A B}{6}-2 A\theta_N \right ) ~~,
\eq
and   $\theta_N$ is a real solution of the auxiliary cubic equation
\bqa
  && \theta_N^3 +a\theta_N +b=0 ~~, \nonumber \\
&& a  \equiv  -\frac{1}{4}\left (-\frac{AC}{4} +\frac{B^2}{12}+D
\right ) ~~, \nonumber \\
&& b \equiv \frac{1}{4} \left (
-\frac{A^2 D}{16}+\frac{A B C}{48}-\frac{B^3}{216}+
\frac{B D}{6}-\frac{C^2}{16} \right ) ~~.
\label{auxiliary-eq}
\eqa
Depending on the signs of $a$ and
\bq
\Delta  \equiv \frac{b^2}{4}+\frac{a^3}{27} ~~,
\label{auxiliary-Delta}
\eq
the expression for $\theta_N$ is constructed using
(14-16) of \cite{GLMP}.

\newpage
\renewcommand{\theequation}{B.\arabic{equation}}
\renewcommand{\thesection}{B.\arabic{section}}
\setcounter{equation}{0}
\setcounter{section}{0}

{\large \bf Appendix B: The projectors and the CP-eigenphases in
terms of the rotation angles and the CP-phases:}\\

The purpose of this Appendix is to clarify the connection between the
present formalism and  the one of \cite{Zerwas1}.
To this aim, the $4 \times 4$ unitary matrix $U_N$ of
(\ref{UN-definition}) is factorized as  \cite{Zerwas1}
\bq
U_N^\dagger=M D ~~ ,\nonumber
\eq
where
\bq
M={\rm Diag} \{e^{i\alpha_1},e^{i \alpha_2}
e^{i \alpha_3},e^{i \alpha_4}\}  ~~ , \label{Majorana-phases}
\eq
is used to define the four Majorana CP phases
satisfying $0 \le \alpha_i < 2 \pi $  \cite{Zerwas1};
while the  D matrix  depends on  6 rotation angles
$\theta_{ij} \in [0,\pi/2],~ (i,j=1-4)$ and 6  CP-phases
$\delta_{ij} \in [0,2 \pi) ,~ (i,j=1-4)$, as \cite{Zerwas1}
\bq
D=R_{34} R_{24} R_{14} R_{23} R_{13} R_{12} ~~,
\eq
with  $R_{ij}$ being  the rotation matrix in the plane $(ij)$.
Thus, \eg
\bqa
R_{12}=\left (  \matrix{ c_{12} & s_{12} e^{-i \delta_{12}} & 0& 0 \cr
 -s_{12}  e^{i \delta_{12}} & c_{12} & 0 & 0 \cr
 0 & 0 & 1 & 0 \cr
 0 & 0 & 0&  1  \cr }\right ) ~~,  ~~\nonumber
\eqa
where\footnote{Note that our convention for $s_{ij}$ slightly differs
from the one used in \cite{Zerwas1}.}
\bq
c_{ij} \equiv \cos \theta_{ij}\ \ , \ \  s_{ij} \equiv
\sin \theta_{ij}\ \ . \ \  t_{ij} \equiv \tan \theta_{ij}~~ .
\label{thetaij-angles}
\eq

In terms of these parameters, the
reduced projector elements $p_{j2}, p_{j3}, p_{j4}$
for the four physical neutralinos, are expressed as
\bqa
p_{12}&=&\kappa_2 t_{12} ~~ , ~~ p_{13}=\kappa_3
\frac{t_{13}}{c_{12}} ~~ , ~~ p_{14}= \kappa_4 \frac{t_{14}}{c_{12}
c_{13}} ~~, \label{p12}  \\
p_{22}&=&\kappa_2 [s_{12} c_{24} s_{13} s_{23}
\xi_{1}-c_{23} c_{24} c_{12}
\xi_2+s_{12}c_{13} s_{14} s_{24}] p_2^{-1}~~,    \\
 p_{23}&=&\kappa_3 [s_{13} s_{14} s_{24}-c_{13} c_{24} s_{23}
\xi_1] p_2^{-1}~~ , \nonumber  \\
p_{24}&=&-\kappa_4  c_{14} s_{24} p_2^{-1}~~,   \\
p_{32}&=&\kappa_2 [s_{12}s_{13} s_{23} s_{24} s_{34}
\xi_1-c_{23} c_{12}
s_{24} s_{34} \xi_2
      -s_{12} c_{23} c_{34} s_{13} \xi_3  \nonumber \\
      &&-c_{34} c_{12} s_{23} \xi_1^* \xi_2
      \xi_3-s_{12} c_{13} c_{24} s_{14} s_{34}] p_3^{-1} ~~,  \\
p_{33}&=&\kappa_3[-c_{13} s_{23} s_{24} s_{34}
\xi_1+c_{13} c_{23} c_{34}
\xi_3-c_{24} s_{13} s_{14} s_{34}] p_3^{-1} ~~,  \\
p_{34}&=& \kappa_4 c_{14} c_{24}
s_{34} p_3^{-1} ~~,  \\
p_{42}&=& \kappa_2 [s_{12}s_{13} s_{23} s_{24} c_{34}
\xi_1-c_{23} c_{12}
s_{24} c_{34} \xi_2
      +s_{12} c_{23} s_{34} s_{13} \xi_3 \nonumber \\
      &&+s_{34} c_{12} s_{23} \xi_1^* \xi_2
      \xi_3-s_{12} c_{13} c_{24} s_{14} c_{34}] p_4^{-1}~~,  \\
p_{43}&=&\kappa_3 [-c_{13} s_{23} s_{24} c_{34}
\xi_1-c_{13} c_{23} s_{34}
\xi_3-c_{24} s_{13} s_{14} c_{34}] p_4^{-1} ~~,   \\
p_{44}&=&\kappa_4 c_{14} c_{24} c_{34}p_4^{-1} ~~, \label{p44}
\eqa
with
\bqa
\kappa_j &=& e^{-i ( \delta_{1j} +\alpha_1-\alpha_{j})} ~~,
\label{kappaj} \\
p_2&=& c_{12} c_{24} s_{13} s_{23}
\xi_{1}+c_{23} c_{24} s_{12} \xi_2+c_{12}c_{13} s_{14} s_{24}
~~, \nonumber \\
p_3&=&c_{12}s_{13} s_{23} s_{24} s_{34}
\xi_1+c_{23} s_{12} s_{24} s_{34} \xi_2
      -c_{12} c_{23} c_{34} s_{13} \xi_3 \nonumber \\
      &&+c_{34} s_{12} s_{23} \xi_1^* \xi_2
      \xi_3-c_{12} c_{13} c_{24} s_{14} s_{34}~~,  \nonumber \\
p_4&=&c_{12}s_{13} s_{23} s_{24}
c_{34} \xi_1+c_{23} s_{12} s_{24} c_{34} \xi_2
      +c_{12} c_{23} s_{34} s_{13} \xi_3 \nonumber \\
      &&-s_{34} s_{12} s_{23} \xi_1^* \xi_2
      \xi_3-c_{12} c_{13} c_{24} s_{14} c_{34}~~,  \label{p4}
\eqa
and
\bq
\xi_1 \equiv e^{i (\delta_{13}-\delta_{23}+\delta_{24}
-\delta_{14})} \ \ , \ \
\xi_2 \equiv e^{i (\delta_{12}+\delta_{24}-\delta_{14})}\ \ , \ \
\xi_3 \equiv e^{i (\delta_{13}+\delta_{34}-\delta_{14})} ~~;
\label{xi}
\eq
while the CP-eigenphases of the four physical
neutralinos defined at (\ref{etaj-def})  are
\bqa
\eta_1&=&e^{2 i \alpha_1}  ~~, \nonumber\\
\eta_2&=&e^{2i(\alpha_{1}+
\delta_{14}-\delta_{24}+{\rm Arg}[p_{14}]
-{\rm Arg}[p_{24}])}~~ , \nonumber\\
\eta_3&=& e^{2i(\alpha_{1}+\delta_{14}-\delta_{34}+{\rm Arg}[p_{14}]
-{\rm Arg}[p_{34}])}~~ , \nonumber\\
\eta_4&=&e^{2i (\alpha_{1}+\delta_{14}+{\rm Arg}[p_{14}]
-{\rm Arg}[p_{44}])}~~. \label{etaj-Zerwas}
\eqa

Eqs.(\ref{p12}-\ref{etaj-Zerwas}) express the parameters of the
present formalism in terms of those of \cite{Zerwas1}.
Conversely, the rotation angles $\theta_{ij}$ and the
 $\alpha_i$ phases of \cite{Zerwas1}
 (compare (\ref{thetaij-angles}, \ref{Majorana-phases})),
 may also be  written in terms of the  projector elements
 and the neutralino CP eigenphases as
\bqa
s_{12}&=&\sqrt{\frac{P_{122}}{1-P_{133}-P_{144}}} \ \ , \ \
s_{13} =\sqrt{\frac{P_{133}}{1-P_{144}}} \ \ , \ \
s_{14}=\sqrt{P_{144}} ~~,  \nonumber \\
s_{24}&=& \sqrt{\frac{P_{244}}{1-P_{144}}}
\ \ , \ \ s_{34}=\sqrt{\frac{P_{344}}{1-P_{144}-P_{244}}} \nonumber \\
s_{23}&=&\sqrt{\frac{(1-P_{144})^2 P_{233}+P_{133} P_{144}
P_{244}+2 (1-P_{144}) Re[P_{134} P_{234}^*]}{(1-P_{133}-P_{144})
(1-P_{144}-P_{244})}}~~ , \label{s12-s23}
\eqa
\bq
\label{CP-phases}
\alpha_i = {\rm Arg}[\sqrt{\eta_i} P_{i14} P_{1i4}]+\tilde{\alpha}_i ~~
, ~~ \delta_{ij}={\rm Arg}[\sqrt{\eta_j \eta_i^*} P_{i14}^*
P_{j14}]+\tilde{\delta}_{ij} ~~,
\eq
with
\bqa
&&\tilde{\alpha}_{1,4} \equiv 0 ~~ , ~~ \tilde{\alpha}_{2,3}
\equiv {\rm Arg} [\xi_{2,3}] ~~ , \nonumber \\
&& \tilde{\delta}_{1i} \equiv
\tilde{\alpha}_i~~ ,~~ \tilde{\delta}_{23}={\rm Arg} [\xi_3 \xi_1^*]
~~ , ~~ \tilde{\delta}_{24}= \tilde{\delta}_{34}=0 ~~ ,
\eqa
where the auxiliary complex numbers $\xi_i$ read:
\bqa
\xi_1&=&\frac{\sqrt{P_{233}}
e^{i {\rm Arg} [P_{134} P_{234}^*]}
+ s_{13} s_{14} s_{24}} {c_{13} c_{24}
s_{23}}~~,  \nonumber\\
\xi_2&=&\frac{s_{12} s_{13} \sqrt{P_{233}}
e^{i {\rm Arg}[P_{134} P_{234}^*]} + c_{13} \sqrt{P_{222}}
 e^{i {\rm Arg} [P_{124} P_{224}^*]}+
s_{12} s_{14} s_{24}} {c_{12} c_{13} c_{23}
c_{24}} ~~ , \nonumber \\
\label{xi3}
\xi_3&=&\frac{s_{24} s_{34} \sqrt{P_{233}}
e^{i {\rm Arg} [P_{134} P_{234}^*]} + c_{24} \sqrt{P_{333}} e^{i
{\rm Arg}[P_{134} P_{334}^*]}+s_{13} s_{14} s_{34}}
{c_{34} c_{13} c_{23} c_{24}} ~~.
\eqa
Notice that  the projectors elements
and the neutralino CP eigenphases appearing above,
may be expressed in terms of the
reduced projector elements through (\ref{P-pj}, \ref{Pj11-pj})
and   (\ref{mchic}).\\

Concerning the  condition
 for CP conservation, we note that the result of
 \cite{Zerwas1} may be expressed as
\bq
\label{CP-cond-Zerwas1}
CP-{\rm conservation} \Leftrightarrow \Bigg \{ \alpha_i= 0 ~~
{\rm or}~~  \frac{\pi}{2} ~~~, ~~~~
\delta_{ij}=0 ~~ {\rm or}~~  \pi ~~ \Bigg \} ,
\eq
which can be immediately verified starting from
 (\ref{basic-cp}) and using (\ref{p12}-\ref{xi},
 \ref{s12-s23}- \ref{xi3}).

 It may also be remarked that since CP conservation according to
(\ref{basic-cp}) is equivalent to the reality of
the reduced projector elements $p_{12}, ~p_{13}, ~p_{14}$,  (which also
implies the reality of $\eta_j$),  the validity
of\footnote{Compare in particular  (\ref{p12}) and (\ref{kappaj}).}
\bqa
\label{CP-cond-Zerwas2}
&&CP-{\rm conservation} \Leftrightarrow
\delta_{1i}=\alpha_i-\alpha_{1} ~~[{\rm mod} ~~ \pi] ~
\eqa
 for  $(i=2,~3, ~ 4)$,  should be  sufficient to   imply the complete
 (\ref{CP-cond-Zerwas1}). The  amusing feature of
 condition (\ref{CP-cond-Zerwas2}), is that it is  equivalent
 to (\ref{CP-cond-Zerwas1}), in spite of the fact that it
 does not directly specify the values of
 any of the $\alpha_j$ or $\delta_{ij}$.

\clearpage
\newpage

\newpage

\clearpage
\newpage

\clearpage
\newpage

\clearpage
\newpage

\end{document}